\documentclass{ws-ijmpa}
\usepackage[super,compress]{cite}
\usepackage{graphicx}
\usepackage{amsmath}

\usepackage{nameref}
\usepackage{hyperref}
\hypersetup{colorlinks,urlcolor=black,citecolor=black,linkcolor=black,filecolor=black}
\usepackage{breakurl}
\usepackage{url}

\usepackage{xcolor}
\usepackage{cancel}

\begin{document}

%
\catchline{}{}{}{}{}
%

\title{
Induced surface and curvature tension equation of state for hadron resonance gas in finite volumes and its relation to morphological 
thermodynamics
}

\author{
	K. A. Bugaev$^{1, 2}$,
	O. V. Vitiuk$^{2, 3}$,
	B. E. Grinyuk$^{1}$, 
	P. P. Panasiuk$^{2}$,
	N. S. Yakovenko$^{2}$,
	E. S. Zherebtsova$^{4, 5}$,
	V. V. Sagun$^{1,6}$, 
	O. I. Ivanytskyi$^{1,7}$, 
	L. V. Bravina$^3$,
	D. B. Blaschke$^{4,7,8}$, 
	S. Kabana$^{9}$, 
	S. V. Kuleshov$^{10}$,
	A. V. Taranenko$^4$,
	E. E. Zabrodin$^{3, 11}$ 
	and 
	G. M. Zinovjev$^1$ 
}

\address{
	$^1$ Bogolyubov Institute for Theoretical Physics, Metrologichna str. 14-B, 03680 Kyiv, Ukraine \\
	$^2$ Department of Physics, Taras Shevchenko National University of Kyiv, 03022 Kyiv, Ukraine \\
	$^3$ Department of Physics, University of Oslo, PB 1048 Blindern, N-0316 Oslo, Norway \\
	$^4$ National Research Nuclear University (MEPhI), Kashirskoe Shosse 31, 115409 Moscow, Russia \\
	$^5$ Institute for Nuclear Research, Russian Academy of Science, 108840 Moscow, Russia \\
	$^6$ CFisUC, Department of Physics, University of Coimbra, 3004-516 Coimbra, Portugal \\
	$^7$ Institute of Theoretical Physics, University of Wroclaw, Max Born Pl. 9, 50-204 Wroclaw, Poland \\
	$^8$ Bogoliubov Laboratory of Theoretical Physics, JINR, Joliot-Curie Str. 6, 141980 Dubna, Russia \\
	$^{9}$ Instituto de Alta Investigaci\'on, Universidad de Tarapac\'a, Casilla 7D, Arica, Chile \\
	$^{10}$ Departamento de Ciencias F\'{\i}sicas, Universidad Andres Bello, Sazi\'e 2212, Santiago, Chile \\
	$^{11}$ Skobeltsyn Institute of Nuclear Physics, Moscow State University, 119899 Moscow, Russia
}

\maketitle

\begin{abstract}
{\bf Abstract.} Here we develop an original approach to investigate the grand canonical partition function of the multicomponent mixtures of Boltzmann particles with hard-core interaction in finite and even small systems of the volumes above 20 fm$^3$. The derived expressions of the induced surface tension equation of state are analyzed in details. It is shown that the metastable states, which can emerge in the finite systems with realistic interaction, appear at very high pressures at which the hadron resonance gas, most probably, is not applicable at all. It is shown how and under what conditions the obtained results for finite systems can be generalized to include into a formalism the equation for curvature tension. The applicability range of the obtained equations of induced surface and curvature tensions for finite systems is discussed and their close relations to the equations of the morphological thermodynamics are established. The hadron resonance gas model on the basis of the obtained advanced equation of state is worked out. Also, this model is applied to analyze the chemical freeze-out of hadrons and light nuclei with the number of (anti-)baryons not exceeding 4, including the most problematic ratios of hyper-triton and its antiparticle. Their multiplicities were measured by the ALICE Collaboration in the central lead-lead collisions at the center-of-mass energy $\sqrt{s_{\rm NN}} =$ 2.76 TeV. 
\keywords{hadron resonance gas model, hard-core repulsion, induced surface and curvature tension, 
finite volumes, morphological thermodynamics}
\end{abstract}

\ccode{PACS numbers: 25.75.-q, 24.10.Pa}

\section{Intorduction}

{During the last years between several successful achievements in developing  the realistic equation of state (EoS) of strongly interacting particles we can emphasize the EoS with the hard-core repulsion formulated on the basis of the induced surface tension (IST) concept \cite{IST1,IST2,IST3}.} In {Refs. \citen{IST1,IST2,IST3}} in the grand canonical ensemble (GCE) formulation, it was shown that the hard-core repulsion between particles generates the surface free energy which can be attributed to the surface tension coefficient. 
The systems studied by the IST EoS include the multicomponent mixtures of nuclei (i.e. with many hard-core radii) of all possible sizes \cite{IST1}, {including highly compressed mixtures of nuclei in compact stars \cite{VetaNS}}, and the mixtures of hadrons with different hard-core radii \cite{IST2,IST3}. More recently the IST concept was successfully applied to modeling the properties of mixtures of hadrons with the light nuclei having the baryonic charge $A \le 4$ and their anti-nuclei \cite{Ref1n,Grinyuk2020,Ref2n}. 

Compared to the traditional versions of the hadron resonance gas model (HRGM) with hard-core repulsion \cite{PBM06} the one based on the IST {approach} \cite{IST2,IST3,Ref1n,Grinyuk2020,Ref2n} employs not only the different hard-core radii of mesons and baryons, but also just three more parameters, namely the hard-core radii of pions $R_{\pi}$=0.15 fm, kaons $R_{K}$=0.395 fm, (anti-)$\Lambda$-hyperons $R_{\Lambda}$=0.085 fm. Note that the hard-core radii of other baryons $R_{b}$=0.365 fm and other mesons $R_{m}$=0.42 fm \cite{IST2,IST3} only slightly differ from {the ones previously obtained} within the Van der Waals approximation \cite{Sagun14}. {These three additional hard-core radii indicate the difference in their interaction from the rest of the particles, that was proven by the achieved} extremely high accuracy in the description of hadronic yields measured in the central nuclear collisions performed in AGS BNL, SPS CERN, RHIC BNL, and LHC CERN. Although the center-of-mass energy in these experiments increase from $\sqrt{s_{NN}} =2.7$ GeV (low AGS BNL collision energy) to $\sqrt{s_{NN}} =2.76$ TeV (LHC CERN energy), i.e. by three orders of magnitude, the {attained} quality of data description $\chi^2/dof \simeq 68.17/60 \simeq 1.13$ \cite{IST2} is not reached by the other formulations of the HRGM  until  now.

Recently the IST concept was generalized further in order to include the effects of curvature tension for quantum hard-spheres \cite{QSTAT2019,LFtrans3}, for two-component classical mixtures of hard spheres and hard-discs \cite{Nazar2019}. The novel EoS based on the concept of induced surface and curvature tensions (ISCT) is able to accurately model the EoS of hard spheres up to the packing fraction $\eta \equiv \sum_k \rho_k V_k \simeq 0.45$ (where $\rho_k$ is the particle number density of the $k$-th sort of particles and $V_k$ is their eigenvolume) and the EoS of hard discs (2-dimensional spheres) up to the packing fraction $\eta \equiv \sum_k \rho_k S_k \simeq 0.7$ (here $S_k$ denotes the eigensurface of the $k$-th sort of particles)\cite{Nazar2019}. This means that the ISCT EoS with a few parameters only is able to reproduce the EoS of classical hard spheres and hard discs in the entire gaseous phase \cite{Mulero} of such systems.

This systematic success is not {accidental}, since, compared to the other {EoSs} with the hard-core repulsion \cite{PBM06,Sagun14,Satarov10,
KAB_Chatter15,TypelHC,Kapusta,Vovchenko,Kadam:2018hdo, Blaschke:2020qrs}, the ISCT EoS has three principal advantages. First of all, it allows one to go far beyond the Van der Waals approximation typically used in {Refs. \citen{PBM06,Sagun14,Satarov10,KAB_Chatter15,TypelHC,Kapusta,Vovchenko}.} Second, {the ISCT is a universal approach} in the sense that it can be applied to {model the} mixtures of classical and quantum particles with hard-core interaction of any range and for any number of different hard-core radii, even for an arbitrarily large number of such radii. Third, from the practical point of view, it is extremely important that the number of equations to be solved for the ISCT EoS is always three and it does not depend on the number of different hard-core radii. All these advantages are also not {accidental} since the ISCT EoS developed in {Refs. \citen{QSTAT2019,LFtrans3,Nazar2019}} is closely related to the concept of {\bf morphological thermodynamics} \cite{MorTer1,MorTer2}. In fact, one can regard the results of {Refs. \citen{QSTAT2019,LFtrans3,Nazar2019}} as a GCE generalization of the morphological thermodynamics approach to the mixtures of quantum particles with hard-core interaction \cite{QSTAT2019,LFtrans3} and the ones of classical hard spheres and hard discs \cite{Nazar2019}.

According to the concept of morphological thermodynamics \cite{MorTer1,MorTer2} the change of free energy of a convex rigid body ${\cal B}$ placed into the fluid away from its critical point and from wetting and drying transitions is solely expressed in terms of system pressure $p$, mean surface tension coefficients $\Sigma$, and two bending rigidities $K$ (or curvature tension coefficient) and $\bar k$: $- \Delta \Omega = p V_{\cal B} + \Sigma S_{\cal B}{\cal B} + K C_{\cal B}+ \bar k X_{\cal B}$, where the coefficients $V_{\cal B}$, $S_{\cal B}$, $C_{\cal B}$ and $X_{\cal B}$ are, respectively, the volume of ${\cal B}$, its surface, mean curvature integrated over the surface area and mean Gaussian curvature also integrated over the surface area. In terms of two local principal curvature radii $R_{c1}$ and {$R_{c2}$} the characteristics $C_{\cal B}$ and $X_{\cal B}$ of ${\cal B}$ can be found explicitly as $C_{\cal B} = \int\limits_{\partial {\cal B}} d^2 r \frac{1}{2} \left[\frac{1}{R_{c1}}+ \frac{1}{R_{c2}}\right] $ and $X_{\cal B} = \int\limits_{\partial {\cal B}} d^2 r \frac{1}{R_{c1} R_{c2}} $ (the Euler characteristic).
 
The concept of morphological thermodynamics \cite{MorTer1,MorTer2} is formulated in the canonical ensemble, and due to this fact it cannot be directly used to describe the properties of strongly interacting matter, since in this case, the number of particles is not conserved, but the fundamental charges (baryonic, electric, strange, etc) are the conserved quantities. Also, it deals with the macroscopic systems, while a truncated version of ISCT EoS, i.e. the IST EoS, was successfully applied \cite{IST1,IST2,IST3,Ref1n,Grinyuk2020,Ref2n} to systems of hadrons and nuclei with a maximal number of particles being essentially below 10$^5$ and the chemical freeze-out volumes being essentially below 10$^5$ fm$^3$. Therefore, it is important both from academic and from practical points of view to understand the reason why ISCT EoS and morphological thermodynamics work so well for rather small systems. Besides, it is of prime importance to find out how and under what conditions the finite volumes of the system can modify the ISCT EoS. These are two major tasks of the present work. 
 
To resolve them we introduce an explicit dependence of the mean hard-core radius $\overline{R}$ that was introduced in {Refs. \citen{Ref1n,Grinyuk2020, Nazar2019}} and was previously evaluated in the thermodynamic limit only. In this work, we analytically evaluate the GCE ensemble partition function for the mixture of hadrons and light nuclei with baryonic charge $A \le 4$ (deuteron, triton, helium-3, hyper-triton, helium-4, and their antiparticles) under assumption $\overline{R}(V)$. The developed formalism allows us to derive the ISCT EoS for the system volumes $V \ge 100$ fm$^3$ and establish a close relation to the morphological thermodynamics for $V \ge 500$ fm$^3$.

{The practical purpose to formulate the HRGM on the basis of ISCT approach is to develop a reliable EoS of hadronic matter which is valid up to the phase transition region(s) expected to exist in QCD matter \cite{Jan_recent}. Moreover, we show how the analysis developed here allows us to conclude that it can be valid in a much wider range of thermodynamic parameters even for small volumes of system and, hence, it can be applied to study the chemical freeze-out analysis in collisions of smaller nuclei, like the calcium and argon.}

The work is organized as follows. In Sect. 2 we define the excluded volumes (second virial coefficients) of particles with hard-core interaction and discuss the major approximation. In Sect. 3, we exactly evaluate the GCE partition function of Boltzmann particles for finite system volumes. Sect. 4 is devoted to the detailed and novel analysis of the singularities of the isobaric partition function of such mixtures for finite systems. In Sect. 5 we analyze the emerged $V$-dependences of the GCE ensemble partition function and self-consistently derive the equation for $\overline{R}(V)$ which for a large system is proportional to the surface tension coefficient $\Sigma$, i.e. $\overline{R}(V) \sim \Sigma$. In this section{,} we also demonstrate how to generalize the obtained results in order to include the curvature tension coefficient $K$ in the developed formalism. The relation of the obtained ISCT EoS for finite systems to the morphological thermodynamics is also discussed in Sect. 5. The multiplicities of hadrons and light nuclei mentioned above that were measured by the ALICE Collaboration in Pb+Pb central collisions at the center-of-mass collision energy $\sqrt{s_{NN}} =2.76$ TeV are analyzed with the {HRGM} based on the ISCT EoS in Sect. 6. Our conclusions are formulated in Sect. 7.

\section{Definition of excluded volumes for a mixture of hadrons and light nuclei}

First, we briefly remind the basic definitions of the classical second virial coefficients and the available volume of the system.
The notations used in these definitions are similar to {Ref. \citen{Ref1n}}. The classical second virial coefficient (excluded volumes per particle) $b_{h_1 h_2}$ of hadrons of radii $R_{h_1}$ and $R_{h_2}$ is defined as
 	\begin{equation}\label{Eq1n}
	b_{h_1 h_2} = b_{h_2 h_1} \equiv \frac{2}{3}\pi(R_{h_1}+R_{h_2})^3 .
	\end{equation}
As it is shown in {Refs. \citen{Grinyuk2020,Ref2n}} the classical second virial coefficient (excluded volume per particle) of a hadron and a nucleus of $A$ baryons can be written as
 \begin{equation}\label{Eq2n}
 b_{Ah} = b_{hA} = A \frac{2}{3}\pi (R_b+R_h)^3\,, 
\end{equation}
where $R_b$ is the hard-core radius of baryons. This result is a reflection of the fact that the light nuclei are roomy clusters. In other words, if one takes the typical hard-core diameter of baryons $2 R_b = 0.73$ fm \cite{IST2,IST3} and the largest hard-core diameter of hadrons (mesons) $0.84$ fm \cite{IST2,IST3}, then the classical distance $L_A$ between the constituents of $A$ baryon nucleus 
is essentially larger than $2(R_b+\max R_h)$. This is true even for the helium-4 nucleus which has the smallest distance between the constituent $L_{He4} \simeq 2.74$ fm due to {the} large binding energy. Hence, it is possible to freely translate the hadron of any type with the hard-core radius $R_h$ around each constituent of the nucleus without touching any other constituent of the very same nucleus \cite{ Ref1n, Grinyuk2020}. 

For further treatment it is useful to introduce the additional degeneracy of nuclei of $A$ baryons $g_{kA}$ and explicitly define the second virial coefficients (\ref{Eq1n}) and (\ref{Eq2n}) as
 \begin{eqnarray}\label{Eq3n}
& b_{k h_l} = g_{kA} \frac{2}{3}\pi (R_k+R_{h_l})^3 = g_{kA} \frac{2}{3}\pi (R_k^3 +3R_k^2R_{h_l}+3R_k R_{h_l}^2+R_{h_l}^3), \qquad \\
 \label{Eq4n}
& {\rm with}~~ g_{kA} \equiv A \delta_{kA} + \delta_{kh} , {\rm and} \quad g_{AA} R_A^n = A R_b^n, \qquad 
\end{eqnarray}
where $\delta_{kA}$ and $\delta_{kh}$ are the usual Kronecker $\delta$ symbols. 

To consider the mixture of hadrons and light nuclei as a gas of Boltzmann particles with hard-core interaction
we, for a moment, neglect the nucleus-nucleus interaction and assume that $b_{A_1 A_2} = 0$. Then the
total excluded volume of such a mixture can be written as 
	\begin{eqnarray} \label{Eq5n}
{V}_{excl}^{tot} = \sum\limits_{k \in h_1, A_1} \sum\limits_{ l \in h_2, A_2} N_k b_{k l} N_l ,
	\end{eqnarray}
where $N_k$ ($N_l$) can be either the number of hadrons of sort $h$ or the number of nuclei of $A$ baryons.
Note that in Eq. (\ref{Eq5n}) the summation runs over all particles and antiparticles since the latter are considered as the independent sorts of particles.

Since the mean number of all hadrons $\sum_h \langle N_h \rangle$ is much larger than the mean number of light nuclei $\langle N_A \rangle$, i.e. $ \langle N_A \rangle \ll \sum_h \langle N_h \rangle$, then for light nuclei we can also write that $ A \langle N_A \rangle \ll \sum_h \langle N_h \rangle$. This result allows one to approximate Eq. (\ref{Eq5n}) as follows 
\begin{eqnarray} \label{Eq6}
{V}_{excl}^{tot} & \simeq & \frac{2}{3}\pi \sum\limits_{k \in h_1, A_1} \sum\limits_{ l \in h_2, A_2} N_k g_{kA_1} (R_k^3 +3R_k^2R_{l}+3R_k R_{l}^2+R_{l}^3)
 N_l g_{l A_2},
\end{eqnarray}
where we substituted the binomial expression (\ref{Eq3n}) and then in Eq. (\ref{Eq6}) we extended the double summation by adding the second degeneracy factor $g_{l A_2}$. In this way, we account for the nucleus-nucleus interaction in a symmetric way which is convenient for further evaluation. It is evident that due to the inequality $ A \langle N_A \rangle \ll \sum_h \langle N_h \rangle$
the approximated Eq. (\ref{Eq6}) is very accurate for A+A collisions \cite{ Ref1n, Ref2n}. 
 
To demonstrate the advantages of the new approach first we consider the IST formulation introduced in \cite{IST1} which is technically simpler. For this purpose{,} we combine the first term in the brackets of Eq. (\ref{Eq6}) with the last term, and the second term with the third one. This allows us to identically rewrite the total excluded volume (\ref{Eq6}) as 
\begin{eqnarray} \label{Eq7n}
{V}_{excl}^{tot} \simeq \frac{4}{3}\pi \sum\limits_{k \in h_1, A_1} \sum\limits_{ l \in h_2, A_2}
 N_k g_{kA_1} (R_k^3 +3R_k^2R_{l}) N_l g_{l A_2}.
\end{eqnarray}
The last results can be used to approximate the mean excluded volume of the system per particle 
\begin{eqnarray} \label{Eq8n}
&&\overline{V}_{excl} = \frac{{V}_{excl}^{tot}}{\sum\limits_{ l \in h, A} N_l } 
 \simeq \frac{{V}_{excl}^{tot}}{\sum\limits_{ l \in h, A} N_l g_{lA} }
 \simeq \sum\limits_{k \in h_1, A_1} N_k g_{kA_1}V_k + \sum\limits_{ k \in h_1, A_1} N_k g_{kA_1} S_k \, \overline{R}
 \,, \quad 
\end{eqnarray}
where $V_k = \frac{4}{3}\pi R_k^3 $ denotes the eigenvolume of the particle of hard-core radius $R_k$, $S_k= 4 \pi R_k^2$ is its eigensurface, while the mean hard-core radius $\overline{R}$ is defined as 
	\begin{equation}
	\label{Eq9n}
	\overline{R} = \sum\limits_{ k \in h, A} N_k g_{kA} R_k \biggl/ \sum\limits_{ l \in h, A} 
 N_l .
	\end{equation}
In what follows it is assumed that the degeneracy $g_{kA}$ acts on the powers of the hard-core radius of particles that appear in $V_k$ and $S_k$ according to the definition (\ref{Eq8n}).

As in our previous works \cite{Ref1n,Nazar2019} here we assume that for sufficiently large, but finite system one can approximate all $ N_k$ values in (\ref{Eq9n}) by their statistical mean values $\left\langle N_k\right\rangle $ and write 
	\begin{equation}
	\label{Eq10n}
	\overline{R} (V) \rightarrow {\sum\limits_{ k \in h, A} \left\langle N_k\right\rangle g_{kA} R_k}\biggl/ {\sum\limits_{l \in h, A}\left\langle N_l\right\rangle} ,
	\end{equation}
where $\left\langle N_l\right\rangle$ will be calculated self-consistently using the grand canonical ensemble (GCE) partition function. In other words, this means that using $\overline{R}$ defined by Eq. (\ref{Eq10n}) we calculate the GCE partition function with the total excluded volume $\overline{V}_{excl}$ (\ref{Eq8n}). The principal difference with our previous works \cite{Ref1n,Nazar2019} is that for any finite size of system the mean hard-core radius of particles $\overline{R}$ defined by Eq. (\ref{Eq10n}) acquires the dependence of system volume $V$. Apparently, the volume dependence of $\overline{R} (V)$ disappears in the thermodynamic limit \cite{Ref1n,Nazar2019}. Therefore, one cannot evaluate the GCE partition for the {volume-dependent} functions $\overline{R} (V)$ and $\overline{V}_{excl}$ with the standard Laplace transform technique to isobaric ensemble as it was done in {Refs. \citen{Ref1n,Nazar2019}}.

Fortunately, this difficulty can be avoided, if we use the Laplace-Fourier technique developed for such cases in {Refs. \citen{LFtrans1,LFtrans2,LFtrans3}.} Hence, in what follows we assume that $\overline{R}$ is a function of temperature $T$, chemical potentials 
$\{\mu_k\}$ and system volume $V$, but {afterward} we will find $\overline{R}$ from the calculated GCE partition. In this way we will be able to self-consistently evaluate the finite volume dependence of $\overline{R}$ in a general way and to find out at what conditions this volume dependence can be safely neglected. Apparently, these conditions will also define the applicability limit of morphological thermodynamics for finite systems. 
	
\section{Laplace-Fourier technique for finite system volumes}
	
Introducing the chemical potential for the $k$-th sort of particles as $\mu_k$, we can write the GCE partition function of the considered mixture of Boltzmann particles as
	\begin{eqnarray}\label{Eq11n}
	& Z_{GCE}(T,\left\lbrace \mu_k \right\rbrace, V) \equiv \sum\limits_{ \left\lbrace { N_k} \right\rbrace }^\infty \left[ \prod\limits_{k \in h, A} \frac{\left[	\phi_k e^{\frac{\mu_k}{T} }(V - \overline{V}_{excl}) \right]^{N_k}}{N_k!} \right] \theta(V - \overline{V}_{excl}) , \\
	& \overline{V}_{excl} = \sum\limits_{ k \in h_1, A_1} N_k g_{kA_1}V_k + \overline{R} (V) \sum\limits_{ k \in h_1, A_1} N_k g_{kA_1} S_k \, ,%
	\label{Eq12n}
	\end{eqnarray}
where {the} thermal density of particles of the $k$-th sort $\phi_k$ is defined as 
\begin{eqnarray}\label{Eq13n}
 \phi_k = g_k \gamma_S^{|s_k|} \int\limits_{M_k^{Th}}^\infty 
\frac{ d m}{N_k (M_k^{Th})} 
\frac{\Gamma_k}{(m-m_{k})^{2}+\Gamma^{2}_{k}/4} \cdot 
\int \frac{d^3 p}{ (2 \pi \hbar)^3 } \exp \left[{\textstyle - \frac{ \sqrt{p^2 + m^2} }{T} }\right] \,.
\end{eqnarray}
Here $g_k$ denotes the degeneracy factor of the $k$-th sort of particle, $m_k$ is its mean mass, $\gamma_S$ denotes its strangeness suppression factor introduced by J. Rafelski \cite{Rafelski}, $|s_k|$ is the number of valence strange quarks and antiquarks in {the corresponding} particle. In Eq. (\ref{Eq13n}) the factor 
\begin{equation}\label{Eq14n}
\displaystyle {N_k (M_k^{Th})} \equiv \int
\limits_{M_k^{Th}}^\infty \frac{d m \, \Gamma_k}{(m-m_{k})^{2}+
\Gamma^{2}_{k}/4} 
\end{equation}
is a normalization constant of the mass attenuation of resonances. Here $M_k^{Th}$ denotes the decay threshold mass of the $k$-th hadronic resonance and $\Gamma_k$ denotes its width. 

Although in Eq.(\ref{Eq13n}) the Breit-Wigner ansatz for the mass attenuation is an approximation which is usually valid for relatively 
narrow resonances only, but it is well known that an expression (\ref{Eq13n}) for the thermal density of unstable particles provides a reasonable accuracy \cite{Blaschke:2013zaa,Kuksa,ResWidth}. 

Apparently, for the stable hadrons and for the stable light nuclei the width $\Gamma_k$ should be set to zero, which results {in} a well-known expression for the thermal density
	\begin{equation}\label{Eq15n}
	\phi_k = g_k \gamma_S^{|s_k|} \int \frac{dp^3}{(2\pi\hbar)^3} \exp \left[{\textstyle - \frac{ \sqrt{p^2 + m^2_k} }{T} }\right] .
	\end{equation}

Note that the Heaviside step function $\theta$ in Eq. (\ref{Eq11n}) is of great importance since it provides the absence of negative values of the available volume $(V - \overline{V}_{excl})$ and leads to the finite number of particles stored in the finite volume of system $V$. However, due to its presence, the evaluation of the GCE partition function (\ref{Eq11n}) is rather complicated. The $\theta$-function constraint on the physics side accounts for the hard-core repulsion in a simple way and, simultaneously, it mathematically ensures that the excluded volumes of particles do not overlap. Unfortunately, the usual Laplace transform with respect to the system volume $V$ cannot be used in this case, since the total excluded volume $\overline{V}_{excl}$ depends on $V$ via the mean hard-core radius $\overline{R} (V)$. Hence our first step is to employ the identity suggested in {Refs. \citen{LFtrans1, LFtrans2}}
\begin{eqnarray}\label{Eq16n}
G(V, V) \equiv \int\limits_{-\infty}^{+\infty} d \xi \, \delta (\xi - V) 
 G(V, \xi ) = \int\limits_{-\infty}^{+\infty} d \xi \, \int\limits_{-\infty}^{+\infty} \frac{d \eta}{2 \pi} \, e^{i\eta (\xi -V)} \, G(V, \xi) \,,
\end{eqnarray}
where the second equality in Eq. (\ref{Eq16n}) is obtained from the first one using the Fourier representation of the Dirac $\delta$-function. The identity (\ref{Eq16n}) transforms any undesired dependence on the system volume $V$ into an exponential one 
\begin{eqnarray}\label{Eq17n}
	& Z_{GCE}(T,\left\lbrace \mu_k \right\rbrace, V ) \equiv 
	\int\limits_{-\infty}^{+\infty} d \xi \int\limits_{-\infty}^{+\infty} \frac{d \eta}{2 \pi} \exp \left[ i\eta ( V - \xi ) \right] \times \nonumber \\
& \times 
\sum\limits_{ \left\lbrace { N_k} \right\rbrace }^\infty \left[ \prod\limits_{k \in h, A} \frac{\left( 	\phi_k \exp \left[ \frac{\mu_k}{T} \right] \left[ V - \overline{V}_{excl}(\xi ) \right]\right)^{N_k}}{N_k!} \right] \theta(V - \overline{V}_{excl}(\xi) )
\,.
\end{eqnarray}
Therefore, now one can evaluate Eq. (\ref{Eq17n}) using the usual Laplace transformation technique with respect to $V$ (for definiteness we consider the spherical systems):
\begin{eqnarray}\label{Eq18n}
&& {\cal Z} (T,\left\lbrace \mu_k \right\rbrace, \lambda) \equiv \int\limits_{0}^{+\infty} d V \, e^{- \lambda V } Z_{GCE}(T,\left\lbrace \mu_k \right\rbrace, V ) = \\
&& \int\limits_{0}^{+\infty} d V \, 	\int\limits_{-\infty}^{+\infty} d \xi \, \int\limits_{-\infty}^{+\infty} \frac{d \eta}{2 \pi} \, 
 \exp \left[ i\eta (V - \xi) 
 - \lambda V \right] \times \, \nonumber \\
\label{Eq19n}
&& \times 
\sum\limits_{ \left\lbrace { N_k} \right\rbrace }^\infty \left[ \prod\limits_{k \in h, A} \frac{\left( 	\phi_k \exp \left[ \frac{\mu_k}{T} \right] \left[ V - \overline{V}_{excl}(\xi ) \right]\right)^{N_k}}{N_k!} \right] \theta(V - \overline{V}_{excl}(\xi) ) \,. 
\end{eqnarray}

To further evaluate Eq. (\ref{Eq19n}) for the isobaric ensemble partition function it is necessary to use the usual trick (for a review see \cite{Bugaev+Reuter}): in order to evaluate the summations in Eq. (\ref{Eq19n}) and to calculate the integral with respect to the variable $V$ one has to change the order of integrals moving the integral with respect to $\lambda$ to the right position and then one has to change the variable from $V$ to $\tilde V = V - \overline{V}_{excl}(\xi)$. Such a change of variable allows one to trivially account for the $\theta$-function constraint
\begin{eqnarray}\label{Eq20n}
	 {\cal Z} (T,\left\lbrace \mu_k \right\rbrace, \lambda ) &=&
	%
 	\int\limits_{-\infty}^{+\infty} d \xi \, \int\limits_{-\infty}^{+\infty} \frac{d \eta}{2 \pi} \, \int\limits_{0}^{+\infty} d \tilde V \, 
 \exp \left[ - (\lambda - i\eta) \tilde V - i\eta \xi) \right] \times \, \nonumber \\
&\times& 
\sum\limits_{ \left\lbrace { N_k} \right\rbrace }^\infty \left[ \prod\limits_{k \in h, A} \frac{\left( 	\phi_k \exp \left[ \frac{\mu_k}{T} \right] \tilde V \right)^{N_k}}{N_k!} \right] \times \nonumber \\
&\times& \exp \left[ - (\lambda - i\eta) \sum\limits_{ l \in h, A } N_l g_{l A} (V_l + S_l \overline{R} (\xi) ) \right] \theta(\tilde V) \,. \quad 
	\end{eqnarray}
Moving in Eq. (\ref{Eq20n}) the terms $N_l g_{l A} (V_l + S_l \overline{R} (\xi) )$ that contain the particle multiplicities $N_l$ to the appropriate places in the product, one can easily make all summations with respect to $N_k$ and obtain the following results 
	\begin{eqnarray}\label{Eq21n}
	&& {\cal Z} (T,\left\lbrace \mu_k \right\rbrace, \lambda)= 
	%
 	\int\limits_{-\infty}^{+\infty} d \xi \, \int\limits_{-\infty}^{+\infty} \frac{d \eta}{2 \pi} \, \int\limits_{0}^{+\infty} d \tilde V \, 
 \exp \left[ - i\eta \xi \right] \times \, \nonumber \\
&& \times 
 \exp \left[ \tilde V \sum\limits_{ k \in h, A } \phi_k \exp \left[ \frac{\mu_k}{T} - (\lambda - i\eta) g_{k A} (V_k + S_k \overline{R} (\xi) ) \right] - (\lambda - i\eta) \tilde V \right] = \quad \\
\label{Eq22n}
&& = \int\limits_{-\infty}^{+\infty} d \xi \, \int\limits_{-\infty}^{+\infty} \frac{d \eta}{2 \pi} \, 
\frac{ \exp \left[ - i\eta \xi \right]}{(\lambda - i\eta) - {\cal F} (\lambda- i\eta, T,\left\lbrace \mu_k \right\rbrace)} \,, 
	\end{eqnarray}
for the isobaric ensemble partition function (\ref{Eq20n}). Note that equality (\ref{Eq22n}) is obtained from Eq. (\ref{Eq21n}) by integration with respect to $\tilde V$, since the $\theta(\tilde V)$-function constraint is now accounted explicitly by the lower limit of $d \tilde V$ integral in Eq. (\ref{Eq21n}).
In Eq. (\ref{Eq22n}) the auxiliary function ${\cal F} (\lambda, T,\left\lbrace \mu_k \right\rbrace)$ is defined as 
\begin{equation}\label{Eq23n}
	 \mathcal{F}(\lambda,T,\left\lbrace \mu_k \right\rbrace) = \sum\limits_{k \in h, A} \phi_k \exp\left[ \frac{\mu_k}{T} - \lambda g_{kA} [V_k + S_k \overline{R}(\xi)] \right] .
\end{equation}		

Eq. (\ref{Eq22n}) is a finite volume generalization of the isobaric partition function found in {Ref. \citen{Ref1n}} for the mixtures of hadrons and light nuclei in the thermodynamic limit. In order to {write down} the equation for the mean hard-core radius $\overline{R}(\xi)$ for finite systems{,} we have to find out the GCE partition function from Eq. (\ref{Eq22n}) for the isobaric partition function. This is done by the inverse Laplace transform with respect to variable $\lambda$
\begin{eqnarray}\label{Eq24n}
	&&Z_{GCE}(T,\left\lbrace \mu_k \right\rbrace, V) = \frac{1}{2\pi i} \int\limits\limits_{\chi - i\infty}^{\chi + i\infty} d\lambda \, e^{\lambda V} \, {\cal Z}(T,\left\lbrace \mu_k \right\rbrace, \lambda) = \\
\label{Eq25n}
 &&= \frac{1}{2\pi i} \int\limits\limits_{\chi - i\infty}^{\chi + i\infty} d\lambda \int\limits_{-\infty}^{+\infty} d \xi \, \int\limits_{-\infty}^{+\infty} \frac{d \eta}{2 \pi} \frac{ \exp \left[ \lambda V - i\eta \xi \right]}{(\lambda - i\eta) - {\cal F} (\lambda- i\eta, T,\left\lbrace \mu_k \right\rbrace)} \, .
	\end{eqnarray}
To evaluate the partition (\ref{Eq25n}) it is, first, necessary to interchange the order of integrals moving the one with respect to 
$\lambda$ to the rightmost position, and then one has to introduce the new variable $\tilde \lambda= \lambda-i\eta$ (more formal details can be found in {Refs. \citen{LFtrans1, Bugaev+Reuter}})
\begin{eqnarray}\label{Eq26n}
	Z_{GCE}(T,\left\lbrace \mu_k \right\rbrace, V) = \frac{1}{2\pi i} \int\limits_{-\infty}^{+\infty} d \xi \int\limits_{-\infty}^{+\infty} \frac{d \eta}{2 \pi} \int\limits\limits_{\tilde \chi - i\infty}^{\tilde \chi + i\infty} d\tilde \lambda \frac{ \exp \left[ \tilde \lambda V + i\eta(V- \xi) \right]}{\tilde \lambda - {\cal F} (\tilde \lambda, T,\left\lbrace \mu_k \right\rbrace)} = 
	\\
	\label{Eq27n}
	= \sum\limits_{\cal N} \exp \left[ \lambda_{\cal N} V \right] \int\limits_{-\infty}^{+\infty} d \xi \int\limits_{-\infty}^{+\infty} \frac{d \eta}{2 \pi} \frac{ \exp \left[ i\eta(V- \xi) \right]}{1 - \frac{\partial {\cal F} (\lambda , T,\left\lbrace \mu_k \right\rbrace)}{\partial \lambda} } \Biggl|_{\lambda = \lambda_{\cal N}} \,,
\end{eqnarray}
where the set of simple poles $\tilde \lambda= \lambda_{\cal N}\,$ (with ${\cal N} = 0, \pm1, \pm 2, ...$) of the isobaric partition in complex $\tilde \lambda$-plane are defined by the equation
\begin{equation}\label{Eq28n}
 \lambda_{\cal N} = \mathcal{F}(\lambda_{\cal N},T,\left\lbrace \mu_k \right\rbrace) 
\Rightarrow 
	\lambda_{\cal N} = \sum\limits_{k \in h, A} \phi_k \exp\left[ \frac{\mu_k- \lambda_{\cal N}\, T g_{kA} [V_k + \overline{R} (\xi) S_k] }{T} \right] \,.
\end{equation}
As usual, to obtain Eq. (\ref{Eq27n}) the integration contour in the complex $\tilde \lambda$-plane in Eq. (\ref{Eq26n}) should be chosen to the right-hand side of the rightmost singularity of all poles $\lambda_{\cal N}$, i.e. $\tilde \chi> \max \{ Re (\lambda_{\cal N}) \}$ (more details can be found in {Ref. \citen{Bugaev+Reuter}}).

It is important to note that the number of hadronic states and light nuclei analyzed here and used in the HRGM is finite \cite{IST2,IST3}, then the sum on the right hand side of Eq. (\ref{Eq28n}) contains a finite number of terms and, hence, as shown in {Ref. \citen{Bugaev+Reuter}}, for finite value of pressure the isobaric partition (\ref{Eq24n}) can have only a finite number of simple poles $\tilde \lambda=\lambda_{\cal N}$ in the complex $\tilde \lambda$-plane. Keeping this in mind, we can evaluate the integral with respect to $\eta$ in Eq. (\ref{Eq27n}) using the fact that neither function $ \mathcal{F}(\lambda,T,\left\lbrace \mu_k \right\rbrace)$ of Eq.(\ref{Eq23n}) nor its derivative $\frac{\partial {\cal F} (\lambda , T,\left\lbrace \mu_k \right\rbrace)}{\partial \lambda}$ entering the right hand side of Eq. (\ref{Eq27n}) depend on the variable $\eta$. Then integrating the Dirac $\delta$-function one can finally write
\begin{eqnarray}\label{Eq29n}
	Z_{GCE}(T,\left\lbrace \mu_k \right\rbrace, V) = \sum\limits_{\cal N} \exp \left[ \lambda_{\cal N} V \right] \int\limits_{-\infty}^{+\infty} d \xi \, \, \frac{ \delta(V- \xi) }{1 - \frac{\partial {\cal F} (\lambda , T,\left\lbrace \mu_k \right\rbrace)}{\partial \lambda} } \Biggl|_{\lambda = \lambda_{\cal N}}
	\nonumber \\
	= \sum\limits_{\cal N} \frac{ \exp \left[ \lambda_{\cal N} V \right] }{1 - \frac{\partial {\cal F} (\lambda , T,\left\lbrace \mu_k \right\rbrace)}{\partial \lambda}} \Biggl|_{\lambda = \lambda_{\cal N}, \,\, \xi= V} 
	\,,
\end{eqnarray}
where in the last step of evaluation we explicitly took into account the fact that $\xi = V$. 

Before deriving the explicit expression for the mean hard-core radius $\overline{R} (V)$ (see Eq. (\ref{Eq10n})) from the GCE partition function (\ref{Eq29n}) it is appropriate to make a few comments here. As usual, the solutions $\lambda_{\cal N}$ of Eq. (\ref{Eq28n}) define the partial (thermal) pressures $p_{\cal N} = T \lambda_{\cal N}$ of ${\cal N}$-th state. In contrast to infinite systems all partial (thermal) pressures $p_{\cal N} = T \lambda_{\cal N}$ of ${\cal N}$-th state with ${\cal N} \ge 1$ have non-zero imaginary parts, 
i.e. ${ \operatorname{Im}} (p_{{\cal N}\ge 1}) \neq 0$, while the state ${\cal N} = 0$ has a real (thermal) pressure $p_0 = T \lambda_{{\cal N}=0}$ \cite{LFtrans1,LFtrans2,Bugaev+Reuter}. The real solution $\lambda_0$ is always the rightmost one \cite{LFtrans1,LFtrans2,Bugaev+Reuter} and this is the only singularity that survives in the thermodynamic limit. The real pole $\lambda_0$ is the stable one, since it generates the largest thermal pressure, while the complex ones correspond to smaller thermal pressures (their real parts, of course) and, hence, for finite systems they are metastable, but due to this very fact, they disappear in the thermodynamic limit. 

From Eq. (\ref{Eq28n}) it is easy to conclude that the solutions $ \lambda_{\cal N \ge 1}$ are coming in complex conjugate pairs. This is {so since} the GCE partition (\ref{Eq29n}) is real by definition. These are general features of the isobaric partition singularities for finite systems in the absence of analog of phase transitions in such systems \cite{LFtrans1,LFtrans2,Bugaev+Reuter}.

\section{Equation for the mean hard-core radius}

A brief analysis of metastable simple poles is present below, but first, we have to obtain the equation for $\overline{R} (V)$. Recalling that the mean multiplicity of {a} particle of sort $k$ is 
	\begin{equation}\label{Eq30n}
	\left\langle N_k \right\rangle \equiv T \frac{\partial}{\partial \mu_k} \ln \left[ Z_{GCE}(T,\left\lbrace \mu_l \right\rbrace, V) \right] \, , 
	\end{equation}
one can identically rewrite Eq. (\ref{Eq10n}) for $\overline{R} (V)$ as 
	\begin{eqnarray}
	\label{Eq31n}
	&&	 \overline{R} (V) 
	= \frac{\sum\limits_{k \in h, A} g_{kA} R_k \frac{\partial}{\partial \mu_k} [ Z_{GCE}(T,\left\lbrace \mu_l \right\rbrace, V)] }{
	\sum\limits_{k \in h, A} \frac{\partial}{\partial \mu_k} [ Z_{GCE}(T,\left\lbrace \mu_l \right\rbrace, V)] } = \\
\label{Eq32n}
&&	 =	
	\frac{\sum\limits_{k \in h, A} g_{kA} R_k \cdot \sum\limits_{\cal N} \frac{ \exp \left[ \lambda_{\cal N} V \right] }{1 - \frac{\partial {\cal F} (\lambda_{\cal N} , T,\left\lbrace \mu_l \right\rbrace)}{\partial \lambda_{\cal N}}} \frac{\partial}{\partial \mu_k} \left[ \lambda_{\cal N} V - \ln (1-\frac{\partial \mathcal{F}(\lambda_{\cal N}, T,\left\lbrace \mu_l \right\rbrace )}{\partial \lambda_{\cal N}} )\right] } 
	{\sum\limits_{k \in h, A} 
	\sum\limits_{\cal N} \frac{ \exp \left[ \lambda_{\cal N} V \right] }{1 - \frac{\partial {\cal F} (\lambda_{\cal N} , T,\left\lbrace \mu_l \right\rbrace)}{\partial \lambda_{\cal N}}} \frac{\partial}{\partial \mu_k} \left[ \lambda_{\cal N} V - \ln (1-\frac{\partial \mathcal{F}(\lambda_{\cal N}, T,\left\lbrace \mu_l \right\rbrace )}{\partial \lambda_{\cal N}} )\right] } \Biggl|_{\xi= V} \,. \quad 
	\end{eqnarray}
To handle the expression above we analyze the metastable simple poles $ \lambda_{{\cal N} \ge 1}$. It should be mentioned that such metastable states do not exist at low pressures (or low particle number densities) and they appear at some threshold conditions which are discussed here. Introducing the real $ \operatorname{Re}_1 \equiv {\operatorname{Re}} ( \lambda_{{\cal N} =1})$ and imaginary $ \operatorname{Im}_1 \equiv {\operatorname{Im}} ( \lambda_{{\cal N} =1}) $ parts of the first complex pole $\lambda_{{\cal N} =1}$, one can explicitly rewrite Eq. (\ref{Eq28n}) as a system 
\begin{eqnarray}\label{Eq33n}
	 \operatorname{Re}_1 = + \sum\limits_{k \in h, A} \phi_k \exp\left[ \frac{\mu_k}{T}- \operatorname{Re}_1 \, g_{kA} V_k^{eff} \right] \cdot \cos\left( g_{kA} V_k^{eff} \operatorname{Im}_1 \right) \,, \\
	\label{Eq34n}
	 \operatorname{Im}_1 = - \sum\limits_{k \in h, A} \phi_k \exp\left[ \frac{\mu_k}{T}- \operatorname{Re}_1 \, g_{kA} V_k^{eff} \right] \cdot \sin\left( g_{kA} V_k^{eff} \operatorname{Im}_1 \right) \,,
\end{eqnarray}
where for the sake of convenience we introduced the short-hand notations for the effective excluded volume $V_k^{eff} \equiv [V_k + \overline{R} (V) S_k]$ of the $k$-th sort of particles. 
 
The right hand side of Eqs. (\ref{Eq33n}) and (\ref{Eq34n}) is an oscillating functions of $ \operatorname{Im}_1$, while at fixed $T$ and $\left\lbrace \mu_k \right\rbrace$ values the amplitude of oscillations is defined solely by $ \operatorname{Re}_1$. These properties lead to two important conclusions. First, since the effective volumes of different particles are different, then all cosine functions staying on the right hand side of Eq. (\ref{Eq33n}) can be equal to a unit, if and only if $ \operatorname{Im}_1 =0$. Therefore, the real simple pole $ \operatorname{Re}(\lambda_0) = \lambda_0$ is always larger than $ \operatorname{Re}_1$ \cite{LFtrans1,LFtrans2,Bugaev+Reuter}, i.e. the pole $\lambda_0$ is always the rightmost one. Apparently, the same is true for other simple poles $\lambda_{{\cal N}> 1}$, i.e. $\lambda_0 > { \operatorname{Re}}(\lambda_{{\cal N}\ge 1})$ and, hence, it is the stable one. 

Second, the right hand side of Eq. (\ref{Eq34n}) is monotonously decreasing function of variable $ \operatorname{Re}_1$ and, hence, if a solution of the system (\ref{Eq33n}) and (\ref{Eq34n}) exists, it should exist at low values of $ \operatorname{Re}_1$. Moreover, inspecting Eq. (\ref{Eq34n}), one can easily conclude that a single solution with $ \operatorname{Im}_1 > 0$ appears, when the straight line $y= \operatorname{Im}_1$ is a tangent line to the curve being the right hand side of Eq. (\ref{Eq34n}). 
In other words, the first metastable simple pole appears, if the solution of system (\ref{Eq33n}) and (\ref{Eq34n}) obeys the equality
\begin{eqnarray}\label{Eq35n}
1 = - \sum\limits_{k \in h, A}g_{kA} V_k^{eff} \phi_k \exp\left[ \frac{\mu_k}{T}- \operatorname{Re}_1 \, g_{kA} V_k^{eff} \right] \cdot \cos\left( g_{kA} V_k^{eff} \operatorname{Im}_1 \right) \,.
\end{eqnarray}
Apparently, the last result is obtained by differentiating Eq. (\ref{Eq34n}) with respect to variable $ \operatorname{Im}_1$.
Now, dividing Eq. (\ref{Eq33n}) by Eq. (\ref{Eq35n}), one can write 
\begin{eqnarray}\label{Eq36n}
 &&\operatorname{Re}_1 = \nonumber \\
 &&= - \frac{\sum\limits_{k \in h, A} \phi_k \exp\left[ \frac{\mu_k}{T}- \operatorname{Re}_1 \, g_{kA} V_k^{eff} \right] \cdot \cos\left[ g_{kA} V_k^{eff} \operatorname{Im}_1 \right] }{\sum\limits_{k \in h, A}
 \hspace*{-1.1mm} g_{kA} V_k^{eff} \phi_k \exp\left[ \frac{\mu_k}{T}- \operatorname{Re}_1 \, g_{kA} V_k^{eff} \right] \cdot \cos\left[ g_{kA} V_k^{eff} \operatorname{Im}_1 \right]} \equiv
-\frac{1}{\overline{V}_k^{eff} }
 \,, \quad \quad 
\end{eqnarray}
where we introduced the averaged effective excluded volume of all particles $\overline{V}_k^{eff}$. Inspection of Eq. (\ref{Eq36n}) shows that the quantity $\overline{V}_k^{eff}$ is positive since it is hard to imagine the conditions of how the numerator of this equation can have a different sign than its denominator if one takes into account that their basic structures are very similar. 

To simplify the justification of the last statement, we note that in each sum of Eq. (\ref{Eq36n}) there is always a dominant term or a group of dominant terms. Apparently, the latter must have the same excluded volume $V_k^{eff}$ and the same baryonic charge. For instance, at high temperatures exceeding the pion mass $m_\pi$ and low values of baryonic chemical potential $\mu_B$, i.e. for $T > m_\pi$ and $T \gg \mu_B$, the pion gas dominates completely. On the other hand, at high values of baryonic chemical potential and sufficiently low temperatures, i.e. for $\mu_B \gg T$ and $m_\pi > T$, the baryons with the smallest excluded volume must dominate. Hence in Eqs. (\ref{Eq33n}), (\ref{Eq34n}), (\ref{Eq35n}) and (\ref{Eq36n}) we can safely keep only the dominant terms which have the same hard-core radii. Under these assumptions{,} Eq. (\ref{Eq36n}) can be written as
\begin{eqnarray}\label{Eq37n}
 \operatorname{Re}_1 \equiv
-\frac{1}{\overline{V}_k^{eff} } \simeq -\frac{1}{V_{domin}^{eff} } < 0 \,, 
\end{eqnarray}
where we introduced the effective excluded volume of the dominant term(s) $V_{domin}^{eff}$. In other words, for the {conditions} mentioned above, i.e. for high temperatures {and high} baryonic chemical potentials, the inequality (\ref{Eq37n}) is valid. Then it is easy to conclude that this inequality should be valid for the intermediate values of $T$ and $\mu_B$. But Eq. (\ref{Eq37n}) is of principal importance, since it shows that the real part of pressure generated by the first complex pole is negative $\operatorname{Re}(p_1) = T { \operatorname{Re}}(\lambda_1) \simeq - \frac{T}{\overline{V}_k^{eff} } < 0$. The latter means that already for the small system volumes $V\sim (10-20) \cdot {\overline{V}_k^{eff} }$ the contribution of the first complex pole to the GCE partition function (\ref{Eq29n}) is simply negligible, since the statistical weight of this state into partition (\ref{Eq29n}) is $\exp\left[ \operatorname{Re}_1 V\right] =\exp\left[- \frac{V}{\overline{V}_k^{eff} }\right] \rightarrow 0$.

From the system (\ref{Eq33n}), (\ref{Eq34n}) and (\ref{Eq35n}) it is easy to conclude that for the first complex simple pole one has to choose $ \overline{V}_k^{eff} \operatorname{Im}_1 = \frac{3}{2}\pi - \delta_1$, where the small phase shift $\delta_1 > 0 $ obeys the inequality $\delta_1 \ll \frac{\pi}{2}$. Apparently, the effective degeneracy factor $g_{kA}$ can be omitted in further evaluations of complex pole properties, since the nuclei are very rare particles and, hence, cannot define the main characteristics of the studied mixtures. 

Substituting this value of $ \operatorname{Im}_1$ into Eq. (\ref{Eq35n}) for the dominant term(s), one can find
\begin{eqnarray}\label{Eq38n}
\sin\left( \delta_1 \right) \simeq \frac{1}{V_{domin}^{eff} \phi_{domin} \exp\left[ \frac{\mu_{domin}}{T} + 1 \right] } \simeq \delta_1 \ll \frac{\pi}{2} \,,
\end{eqnarray}
where we used the result (\ref{Eq37n}) for $ \operatorname{Re}_1$ and the smallness of the phase shift $\delta_1$ to establish the inequality. Recalling now the hard-core radii of hadrons, one immediately concludes that $V_{domin}^{eff} \le 1$ fm$^3$ and, hence, inequality (\ref{Eq38n}) can be also cast as 
\begin{eqnarray}\label{Eq39n}
 \phi_{domin} \exp \left[ \frac{\mu_{domin}}{T} + 1 \right] \gg \frac{2}{\pi V_{domin}^{eff} } > 0.6 ~ {\rm fm}^{-3} \,,
\end{eqnarray}
where $\mu_{domin}$ denotes the baryonic chemical potential of {the} dominant term(s). Note that the quantity {$\frac{2}{\pi V_{domin}^{eff} } > 0.6 ~{\rm fm}^{-3}$} is rather large for the HRGM, since such a particle number density is about the four times of the ground particle number density of nuclei. Thus, the first complex pole of the isobaric partition may appear in rather dense systems, i.e. for very high values of $T$ and/or $\frac{\mu_{domin}}{T}$. Clearly, for the values of $T$ and $\mu_{domin}$ that are not obeying the inequality (\ref{Eq39n}) there exists only the real simple pole $\lambda_0$.

It is appropriate to note here that the results of Eqs. (\ref{Eq37n}) and (\ref{Eq39n}) are consistent with the common knowledge that the ideal gas of Boltzmann particles is stable at any pressure, any density, and any finite (but not vanishing!) volume. Indeed, taking the limit $V_{domin}^{eff} \rightarrow 0$ in Eqs. (\ref{Eq37n}) and (\ref{Eq39n}), one finds that the pressure of the first complex pole of an ideal gas goes to $- \infty$, and such metastable state can appear at unlimitedly large values of temperature or chemical potential. Evidently, such states are suppressed even at finite volumes and, hence, they cannot exist. 

Combining Eqs. (\ref{Eq38n}) and (\ref{Eq34n}) for the dominant term(s), one can get a useful estimate for the complex part $ \operatorname{Im}_1$ of the considered simple pole
\begin{eqnarray}\label{Eq40n}
 \operatorname{Im}_1 \simeq \frac{1}{V_{domin}^{eff} {\rm tg}(\delta_1)} \simeq \frac{1}{V_{domin}^{eff} \delta_1} \,,
\end{eqnarray}
which goes to $+\infty$ in the limit $V_{domin}^{eff} \rightarrow 0$. 

Therefore, from the analysis above we conclude that the contributions of all complex states in the GCE partition function (\ref{Eq29n}) and in the definition of mean hard-core radius $\overline{R} (V)$ (\ref{Eq31n}) can be safely neglected even for the small system volumes $V > (10-20) \cdot {\max\{ {V}_k^{eff} \} } \simeq 10-20$ fm$^3$ and the particle number densities below $\frac{2}{\pi V_{domin}^{eff} }$. Under these assumptions Eq. (\ref{Eq31n}) can be greatly simplified to 
	\begin{eqnarray}
	\label{Eq41n}
	&&	 \overline{R} (V) =
	\frac{\sum\limits_{k \in h, A} g_{kA} R_k 	\frac{\partial}{\partial \mu_k} \left[ \lambda_0 - \frac{1}{ V}\ln (1-\frac{\partial \mathcal{F}(\lambda_0, T,\left\lbrace \mu_l \right\rbrace )}{\partial \lambda_0} )\right] } 
	{\sum\limits_{k \in h, A} 
	 \frac{\partial}{\partial \mu_k} \left[ \lambda_0 - \frac{1}{ V}\ln (1-\frac{\partial \mathcal{F}(\lambda_0, T,\left\lbrace \mu_l \right\rbrace )}{\partial \lambda_0} )\right] } \Biggl|_{\xi= V} \,.
	\end{eqnarray}
This equation, apparently, has to be supplemented by the real solution $ \lambda_0$ of Eq. (\ref{Eq28n}). In the next section,
 we analyze Eq. (\ref{Eq41n}) in details and elucidate its relation to the morphological thermodynamics.

\section{Inclusion of the curvature tension and further generalization}

Using the explicit expression for the function $\mathcal{F}(\lambda,T,\left\lbrace \mu_k \right\rbrace)$ of Eq. (\ref{Eq23n}) and Eq. (\ref{Eq28n}), one can find the following derivatives
	\begin{eqnarray}\label{Eq42n}
& \frac{\partial^2 \mathcal{F}(\lambda_0, T,\left\lbrace \mu_l \right\rbrace )}{\partial \mu_k \partial \lambda_0} = - \frac{ g_{kA} V_k^{eff}}{T} \phi_k \exp\left[ \frac{\mu_k}{T} - \lambda_0 g_{kA} V_k^{eff} ] \right] + \lambda_0 \frac{\partial \lambda_0 }{\partial \mu_k} \langle \left(g_{lA} V_l^{eff} \right)^2 \rangle 
	\,, \quad \\
\label{Eq43n}
 & {\rm with}~~ \langle \left(g_{lA} V_l^{eff} \right)^2 \rangle \equiv \lambda_0^{-1} \sum\limits_{l \in h, A} \left(g_{lA} V_l^{eff} \right)^2 \phi_l \exp\left[ \frac{\mu_l}{T} - \lambda_0 g_{lA} V_l^{eff} \right] 
 \, , \quad \\
\label{Eq44n}
 & \frac{\partial \lambda_0 }{\partial \mu_k} = \frac{\phi_k \exp\left[ \frac{\mu_k}{T} - \lambda_0 g_{kA} V_k^{eff} \right] }{T \left[ 1-\frac{\partial \mathcal{F}(\lambda_0, T,\left\lbrace \mu_l \right\rbrace )}{\partial \lambda_0} \right] } \Biggl|_{\xi= V} \,,
	\end{eqnarray}
regarding the quantity $\overline{R} (V) $ as a constant of $\{\mu_k\}$. Substituting these results into Eq. (\ref{Eq41n}), one gets 
\begin{eqnarray}
\label{Eq45n}
	&&	 \overline{R} (V) = 
	\nonumber \\
	&&	 = \frac{\sum\limits_{k \in h, A} \hspace*{-1.1mm} g_{kA} R_k 
	\phi_k \exp\left[ \frac{\mu_k}{T} - \lambda_0 g_{kA} V_k^{eff} \right] \hspace*{-1.1mm}	
	\left( 1 - \frac{g_{kA} V_k^{eff}}{V} + \frac{\lambda_0 \langle \left(g_{lA} V_l^{eff} \right)^2 \rangle}{V \left[1-\frac{\partial \mathcal{F}(\lambda_0, T,\left\lbrace \mu_l \right\rbrace )}{\partial \lambda_0} \right] }
	\right) } 
	{\sum\limits_{k \in h, A} \hspace*{-1.1mm} \phi_k \exp\left[ \frac{\mu_k}{T} - \lambda_0 g_{kA} V_k^{eff} \right]	\hspace*{-1.1mm}
	\left( 1 - \frac{g_{kA} V_k^{eff}}{V} + \frac{\lambda_0 \langle \left(g_{lA} V_l^{eff} \right)^2 \rangle}{V \left[1-\frac{\partial \mathcal{F}(\lambda_0, T,\left\lbrace \mu_l \right\rbrace )}{\partial \lambda_0} \right] }
	\right) } \Biggl|_{\xi= V} \,. \nonumber
	 \\
\end{eqnarray}
It is easy to see that the terms which are proportional to the mean value $\langle \left(g_{lA} V_l^{eff} \right)^2 \rangle$ are small in any practical applications of the HRGM. Indeed, using the estimate $\langle \left(g_{lA} V_l^{eff} \right)^2 \rangle \le \max\{ \left( V_l^{eff} \right)^2 \} \simeq 1$ fm$^6$, one can show the equivalence of the following conditions 
\begin{eqnarray}\label{Eq46n}
\frac{p\langle \left(g_{lA} V_l^{eff} \right)^2 \rangle}{T V } \ll 1 \Leftrightarrow p \ll \frac{T V }{1\, {\rm fm}^6} \,. 
\end{eqnarray}
Noting that even for the low temperature $T=10$ MeV and the system volume $V = 100$ fm$^3$ not only the pressure $\frac{T V }{1\, {\rm fm}^6} \simeq 10^3$ MeV$\cdot$fm$^{-3}$, but its tenth part are too large for the hadron gas and, hence, for the HRGM. Hence for $T > 10$ MeV and $V > 100$ fm$^3$ all terms staying in the largest brackets in Eq. (\ref{Eq45n}) are small corrections compared to a unit. Therefore, all these corrections can be moved into the exponential function and, hence, one can write
\begin{eqnarray}
	\label{Eq47n}
	&&	 \overline{R} (V) \simeq 
\frac{\sum\limits_{k \in h, A} g_{kA} R_k 
	\phi_k \exp\left[ \frac{\mu_k}{T} - \left( \lambda_0 + \frac{1}{V} \right)g_{kA} V_k^{eff} \right]	
	 }
	{\sum\limits_{k \in h, A} \phi_k \exp\left[ \frac{\mu_k}{T} - \left( \lambda_0 + \frac{1}{V} \right) g_{kA} V_k^{eff} \right]	
 }
 \,.
	\end{eqnarray}
It is remarkable that in Eq. (\ref{Eq47n}) the both corrections proportional to the mean value $\langle \left(g_{lA} V_l^{eff} \right)^2 \rangle$ cancelled each other. 

From the present analysis one can deduce that Eq. (\ref{Eq47n}) can be safely used for $T > 10$ MeV and systems with the volume 
of about $V \simeq 100$ fm$^3$ (the system radius, in this case, is about 2.9 fm only!). In fact, Eq. (\ref{Eq47n}) can provide about 10 \% accuracy, if it is used for smaller system volumes, namely for 50 fm$^3 \le V < 100$ fm$^3$, while for even smaller volumes 50 fm$^3$ $> V > 20$ fm$^3$ one has to use a more complicated Eq. (\ref{Eq45n}). However, in all the cases discussed in the present paragraph the Eq. (\ref{Eq45n}) and, consequently, Eq. (\ref{Eq47n}) differ from the ones of morphological thermodynamics. 

This is so, due to the explicit presence of system volume $V$ in the denominator in Eq. (\ref{Eq47n}), this denominator does not match the quantity $\lambda_0 = \frac{p_0}{T}$ defined by Eq. (\ref{Eq28n}). Therefore, to use Eqs. (\ref{Eq45n}) and (\ref{Eq47n}) one has to modify the equations of morphological thermodynamics. Fortunately, as shown in Ref.  \cite{Oliinych2013} the minimal chemical freeze-out volume found for collisions of heavy ions is about $\min \{ V \} \simeq 500$ fm$^3$  and, therefore, to study the chemical freeze-out process in collisions of gold-gold or lead-lead nuclei the discussed corrections in the largest brackets of Eq. (\ref{Eq45n}) become negligible and one can omit them. Therefore, for $V \ge 500$ fm$^3$ instead of Eq. (\ref{Eq47n}) from Eq. (\ref{Eq45n}) one obtains
	\begin{eqnarray}
	\label{Eq48n}
	&&	 \overline{R} = 
\frac{\sum\limits_{k \in h, A} g_{kA} R_k 
	\phi_k \exp\left[ \frac{\mu_k}{T} - g_{kA} [ V_k + \overline{R} S_k ] \frac{p_0}{T} \right]	
	 }
	{\sum\limits_{k \in h, A} \phi_k \exp\left[ \frac{\mu_k}{T} - g_{kA} [ V_k + \overline{R} S_k ] \frac{p_0}{T} \right]	
 } \,, 
 	\end{eqnarray}
where the mean hard-core radius $\overline{R}$ does not depend on the system volume anymore. In Eq. (\ref{Eq48n}) we used the definition of effective excluded volume $V_k^{eff}$ for the volume independent mean hard-core radius $\overline{R}$.

Now instead of the quantity $\overline{R}$ one can introduce the traditional quantity of the morphological thermodynamics, namely the surface tension coefficient
	\begin{eqnarray}
	\label{Eq49n}
&&	 \Sigma \equiv \overline{R} p_0 = T \sum\limits_{k \in h, A} g_{kA} R_k 
	\phi_k \exp\left[ \frac{\mu_k - g_{kA} [ V_k p_0 + S_k \Sigma]}{T} \right] \,.
	\end{eqnarray}
The system thermal pressure can be written now in a similar way 
	\begin{eqnarray}
	\label{Eq50n}
&&	 p_0 = T \sum\limits_{k \in h, A} 
	\phi_k \exp\left[ \frac{\mu_k - g_{kA} [ V_k p_0 + S_k \Sigma]}{T} \right] \,.
	\end{eqnarray}
It is evident that one can repeat all the steps used in deriving Eqs. (\ref{Eq49n}) and (\ref{Eq50n}) for a different parameterization of the total excluded volume (\ref{Eq6}), namely without combining the second term with the third one in Eq. (\ref{Eq6}). In this case, the mean excluded volume of the system per particle can be written as (for more details see {Refs. \citen{Ref1n, Nazar2019}})
\begin{eqnarray} \label{Eq51n}
&& \overline{V}_{excl} = {V}_{excl}^{tot} \biggl/ \sum\limits_{ l \in h, A} 
 N_l g_{lA} \simeq \sum\limits_{ l \in h, A} 
N_k g_{kA_1} (V_k + A S_k \, \overline{R} + B C_k \overline{R^2} ) \,,
\end{eqnarray}
where in addition to the mean hard-core radius $\overline{R}(V)$ we introduce the volume-dependent hard-core radius squared $\overline{R^2} (V)$ 
	\begin{equation}
	\label{Eq52n}
	\overline{R^2} (V) = {\sum\limits_{ k \in h, A} \left\langle N_k\right\rangle g_{kA} R_k^2}\biggl/ {\sum\limits_{l \in h, A}\left\langle N_l\right\rangle} \,.
	\end{equation}
In Eq. (\ref{Eq51n}) the quantity $C_k = 4 \pi R_k$ is the double eigenperimeter of the particles of $k$-th sort, while the coefficients $A= \frac{1}{2}$ and $B= \frac{1}{2}$ later on will be considered as the adjustable parameters. 

After repeating the derivation presented above for the effective excluded volume defined as $V_k^{eff}=V_k + A S_k \, \overline{R} + B C_k \overline{R^2}$, one again obtains Eq. (\ref{Eq45n}) for $\overline{R} (V)$ under the validity of Eqs. (\ref{Eq37n}) and {(\ref{Eq39n})}, but with the newly defined value of $V_k^{eff}$. In addition one also gets a similar equation for $\overline{R^2} (V)$
	\begin{eqnarray}
	\label{Eq53m}
		 \overline{R^2} (V) = && \nonumber \\
	&	\hspace*{-3.3mm} = \frac{\sum\limits_{k \in h, A} g_{kA} R_k^2 
	\phi_k \exp\left[ \frac{\mu_k}{T} - \lambda_0 g_{kA} V_k^{eff} \right]	
	\left( 1 - \frac{g_{kA} V_k^{eff}}{V} + \frac{\lambda_0 \langle \left(g_{lA} V_l^{eff} \right)^2 \rangle}{V \left[1-\frac{\partial \mathcal{F}(\lambda_0, T,\left\lbrace \mu_l \right\rbrace )}{\partial \lambda_0} \right] }
	\right) } 
	{\sum\limits_{k \in h, A} \phi_k \exp\left[ \frac{\mu_k}{T} - \lambda_0 g_{kA} V_k^{eff} \right]	
	\left( 1 - \frac{g_{kA} V_k^{eff}}{V} + \frac{\lambda_0 \langle \left(g_{lA} V_l^{eff} \right)^2 \rangle}{V \left[1-\frac{\partial \mathcal{F}(\lambda_0, T,\left\lbrace \mu_l \right\rbrace )}{\partial \lambda_0} \right] }
	\right) 
 } \Biggl|_{\xi= V} \,. \nonumber \\
	\end{eqnarray}

Evidently, for large system volumes $V \ge 500$ fm$^3$ one can further generalize Eqs. for $\overline{R}$ and $\overline{R^2}$ and arrive  at the following system of equations for the ISCT EoS
	\begin{eqnarray}
	\label{Eq53n}
&&	 p = T \sum\limits_{k \in h, A} 
	\phi_k \exp\left[ \frac{\mu_k - g_{kA} [ V_k p + S_k \Sigma + C_k K]}{T} \right] \,,\\
	\label{Eq54n}
&&	 \Sigma \equiv A \overline{R} \, p = A T \sum\limits_{k \in h, A} g_{kA} R_k 
	\phi_k \exp\left[ \frac{\mu_k - g_{kA} [ V_k p + \alpha_k S_k \Sigma + C_k K]}{T} \right] \,,\\
	\label{Eq55n}
&&	 K \equiv B \overline{R^2}\, p = B T \sum\limits_{k \in h, A} g_{kA} R_k^2 
	\phi_k \exp\left[ \frac{\mu_k - g_{kA} [ V_k p + \alpha_k S_k \Sigma + \beta_k C_k K]}{T} \right] \,, \quad \quad 
	\end{eqnarray}
where in the equations for the surface tension $\Sigma$ and curvature tension $K$ coefficients we introduced {$\{ \alpha_k \}$ and $\{ \beta_k \}$, respectively}. These coefficients are introduced according to the concept of the ISCT \cite{Ref1n, Nazar2019,LFtrans3,QSTAT2019}.
In {Ref. \citen{Nazar2019}} one can find various examples demonstrating that for $\alpha_k > 1$ and $\beta_k > 1$ the system (\ref{Eq53n})-(\ref{Eq55n}), and its generalizations for classical and quantum hard spheres \cite{LFtrans3}, 
{as it is mentioned in the Introduction, are able to provide an accurate description of the EoS properties of hard spheres up to the packing fraction $\eta \equiv \sum_k \rho_k V_k \simeq 0.45$ and even the one of hard discs up to the packing fraction $\eta \equiv \sum_k \rho_k S_k \simeq 0.7$.} In other words, in the GCE formulation the ISCT EoS with a few parameters only is able to reproduce the EoS of classical hard spheres and hard discs in the entire gaseous phase \cite{Nazar2019, Mulero} of such systems. Therefore, one can consider the system (\ref{Eq53n})-(\ref{Eq55n}) as the GCE ensemble formulation of the morphological thermodynamics \cite{MorTer1,MorTer2} generalized to the mixtures of hadrons and light nuclei with any (even arbitrary large) number of different hard-core radii $\{R_k\}$. An apparent reason for such a success is the fact that for $\alpha_k > 1$ and $\beta_k > 1$ the effective excluded volume of particles $V_k^{eff} \equiv [p V_k + \Sigma S_k + K C_k] p^{-1}$ is a decreasing function of system pressure \cite{Ref1n, Nazar2019,LFtrans3,QSTAT2019}, i.e. at high pressures the ratios $\frac{\Sigma}{p}$ and $\frac{K}{p}$ tend to vanish and, thus, the effective excluded volumes of particles are compressible to some extent. 
 
It is interesting to note that the system (\ref{Eq53n})-(\ref{Eq55n}) does not contain the term of mean Gaussian curvature $X_k$ of $k$-th sort of particles, which exists in the morphological thermodynamics formulation in the canonical ensemble. The apparent mathematical reason for its absence in the GCE formulation is that the discussed excluded volumes of hard spheres (\ref{Eq1n})-(\ref{Eq7n}) do not contain it. One could, of course, write a separate equation for its conjugate variable $\bar k$, but it seems that this is not necessary, since the whole term $\bar k X_k$ has a dimension of energy and, hence, it can be absorbed in the chemical potential $\mu_k$ of $k$-th sort of particles, which is an independent variable of the GCE ensemble.

This is so, since the quantities $\overline{R} (V)$ and $\overline{R^2} (V)$ are real by construction and, hence, one has to solve only the equation for pressure (\ref{Eq53n}) for the volume dependent functions $\Sigma (V)$ and $K(V)$. Furthermore, in our present analysis we were dealing with the effective excluded volume of particle $V_k^{eff}$ only, and not with the quantities $V_k$ and $S_k \overline{R} (V)$ separately. Due to this important property one can easily generalize the results of Eqs. (\ref{Eq33n})-(\ref{Eq36n}) to the effective volume $V_k^{eff} \equiv g_{kA} [V_k + \overline{R} (V) S_k + \overline{R^2} (V)C_k]$ that is a decreasing function of system pressure. 

As a result, at high pressures Eqs. (\ref{Eq37n}) and (\ref{Eq39n}) should be modified as follows
\begin{eqnarray}\label{Eq56n}
 \operatorname{Re}_1 \simeq -\frac{1}{V_k^{domin} } < 0 \,, \quad 
\phi_{domin} \exp \left[ \frac{\mu_{domin}}{T} + 1 \right] \gg \frac{2}{\pi V_k^{domin} } > 2.4~ {\rm fm}^{-3}
\,,
\end{eqnarray}
where instead of effective excluded volume of the dominant term(s) we have to use the eigenvolume of the dominant term(s) which is four times smaller than the excluded volume, i.e. $\max\{ V_k^{domin}\} \simeq 0.25$ fm$^{3}$. These results demonstrate us that at high pressures the metastable states of isobaric partitions defined by the ISCT EoS will be even stronger suppressed for finite volumes and, hence, the ISCT EoS (\ref{Eq53n})-(\ref{Eq55n}) could be reliably used at very high particle number densities, namely up to 10 values of the ground particle number density of nuclei, but, most probably, the hadronic degrees of freedom are irrelevant at so large particle number densities. 

If, however, one needs to apply the ISCT EoS for the system volumes $V$ about 100-500 fm$^{3}$, then, instead of Eqs. (\ref{Eq54n}) and (\ref{Eq55n}), it is more reliable to use the following analogs of Eq. (\ref{Eq47n}) for $ \overline{R} (V)$ and $ \overline{R^2} (V)$
\begin{eqnarray}
	\label{Eq58n}
&&	 \overline{R} (V) \simeq \nonumber \\
&&	 \simeq 
\frac{\sum\limits_{k \in h, A} g_{kA} R_k 
	\phi_k \exp\left[ \frac{\mu_k}{T} - \left( \frac{p}{T}+ \frac{1}{V} \right)g_{kA}
	 [V_k + A S_k \alpha_k \, \overline{R}(V) + B C_k \overline{R^2}(V)] \right]	
	 }
	{\sum\limits_{k \in h, A} \phi_k \exp\left[ \frac{\mu_k}{T} - \left( \frac{p}{T}+ \frac{1}{V} \right)g_{kA}
	 [V_k + A S_k \, \overline{R}(V) + B C_k \overline{R^2}(V)]\right]	
 }
 \,, \quad \nonumber \\
\end{eqnarray}
\begin{eqnarray}
\label{Eq59n}
&&	 \overline{R^2} (V) \simeq \nonumber \\
&&	 \simeq 
\frac{\sum\limits_{k \in h, A} g_{kA} R_k^2 
	\phi_k \exp\left[ \frac{\mu_k}{T} - \left( \frac{p}{T}+ \frac{1}{V} \right)g_{kA}
	 [V_k + A S_k \alpha_k \, \overline{R}(V) + B C_k \beta_k \overline{R^2}(V)] \right]	
	 }
	{\sum\limits_{k \in h, A} \phi_k \exp\left[ \frac{\mu_k}{T} - \left( \frac{p}{T}+ \frac{1}{V} \right)g_{kA}
	 [V_k + A S_k \, \overline{R}(V) + B C_k \overline{R^2}(V)]\right]	
 }
 \,, \quad \nonumber \\
	\end{eqnarray}
	which have to be supplemented by the equation for pressure, written in the same terms 
	\begin{eqnarray}
	\label{Eq60n}
&&	 p = T \sum\limits_{k \in h, A} 
	\phi_k \exp\left[ \frac{\mu_k - p\, g_{kA}
	 [V_k + A S_k \, \overline{R}(V) + B C_k \overline{R^2}(V)]}{T} \right] \,.
	\end{eqnarray}
The system (\ref{Eq58n})-(\ref{Eq60n}) is the ISCT EoS formulated for small volumes. For one and two-component systems of hard-spheres and hard-discs the coefficients $\{\alpha_k\}$, $\{\beta_k\}$, $A$ and $B$ are given in \cite{Nazar2019}, while for the other multicomponent mixtures in 3-dimensional space they can be found from fitting the pressure of such mixtures with the help of the well-known multicomponent Mansoori-Carnahan-Starling-Leland (MCSL) EoS suggested in {Ref. \citen{MCSLeos}}. Unfortunately, for the 2-dimensional hard-spheres, i.e. for the hard discs, only the EoS at high packing fractions are well documented only for two-component mixtures \cite{Santoseos}.

\section{Chemical freeze-out of light nuclei}

In this work we apply the newly developed HRGM model based on Eqs. (\ref{Eq53n})-(\ref{Eq55n}) to describe the ALICE data on light nuclear cluster yields \cite{KAB_Ref1a,KAB_Ref1b,KAB_Ref1c} measured at the center-of-mass collision energy $\sqrt{s_{NN}} =2.76$ TeV. Since all the details of the fitting procedure of the hadronic data are well documented in {Refs. \citen{Ref1n,IST2,IST3}}, we do not discuss them here. Instead we are concentrating on the analysis of the light nuclei yields data with the setup of ISCT EoS taken from {Ref. \citen{Nazar2019}}. According to the values of the hard-core radii of hadrons we divide them into two groups: the particles with small hard-core radii, i.e. the pions and (anti-)$\Lambda$-hyperons with $\alpha_\pi = \alpha_\Lambda = 1.007$ and $\beta_\pi = \beta_\Lambda = 1.862$, and the ones with large hard-core radii, i.e. other baryons, {kaons}, other mesons and nuclei with $\alpha_b = \alpha_m = 1.050$ and $\beta_b = \beta_m = 2.084$. The parameters $A= 0.443$ and $B=0.63$ are taken from the Table I of {Ref. \citen{Nazar2019}}. From these parameters one concludes that the particles with smaller hard-core radii are suppressed less and this suppression mainly affects the curvature tension coefficient $K$. 
 
The ISCT EoS results are compared here with the HRGM based on the IST EoS developed in {Refs. \citen{Ref1n,Grinyuk2020,Ref2n}}. The latter corresponds to the system (\ref{Eq53n})-(\ref{Eq55n}) with the following parameters: $\alpha_k=1.245$, $\beta_k = 1$ for all particles, while A=1 and B=0. 

Following our ideology presented in details in {Refs. \citen{Ref1n,Grinyuk2020}}, we would like to verify two different scenarios of the chemical freeze-out (CFO) of light nuclei: scenario I or a single CFO scenario, when it occurs together with the CFO of hadrons, and scenario II or a separate CFO of nuclei from the hadrons. Furthermore, in our previous studies, we never included into the fit the problematic ratios 
\begin{equation}\label{eq_18}
	S_3 = \frac{{}^{3}_{\Lambda}{\rm H}}{{}^{3}{\rm He}} \times \frac{p}{\Lambda}, \quad \overline{S}_3 = \frac{{}^{3}_{\overline{\Lambda}}\overline{{\rm H}}}{{}^{3}\overline{{\rm He}}} \times \frac{\overline{p}}{\overline{\Lambda}},
\end{equation} 
which include the hyper-triton $^{3}_{\Lambda}{\rm H}$ and its antiparticle yields. Recently these ratios were measured by the ALICE Collaboration \cite{KAB_Ref1c}. They are of special interest to us since they include the hyper-triton and $\Lambda$-hyperon yields. According to our previous works the hard-core radius of (anti-)$\Lambda$-hyperons is essentially smaller $R_{\Lambda}$=0.085 fm \cite{Sagun14,Ref1n,Ref2n}, than the one of other baryons $R_{b}$=0.365 fm. Hence, the excluded volume of the hyper-triton nucleus almost coincides with one of the deuteron \cite{Ref2n}. Therefore, it is of great interest {for} us to see, if a more advanced formulation of the HRGM based on the ISCT EoS can provide us with some new results. 

In the scenario I (single CFO) without fitting the $S_3$ and $\bar S_3$ ratios, one has 10 experimental data points for hadronic ratios, 8 yields of light nuclear clusters and 2 fitting parameters (same CFO temperatures of hadrons $T_h$ and nuclei $T_A$, i.e. $T_h=T_A$, and the CFO volume $V_A$ of nuclei), while all chemical potentials are set to zero \cite{IST2,IST3,Ref1n}. If the $S_3$ and $\bar S_3$ ratios are included, then one has 20 data points to fit with two parameters. Scenario I is the least probable scenario, as it was found in Refs. \citen{Ref1n, Ref2n} for the IST EoS, since it always leads to an essentially larger value of $\chi^2_{tot}/dof$ at its minimum compared to scenario II. Our present study confirms this finding and, hence, we mainly concentrate on the analysis of scenario II.
 \begin{figure}[th]
	\centerline{\includegraphics[width=0.5\columnwidth]{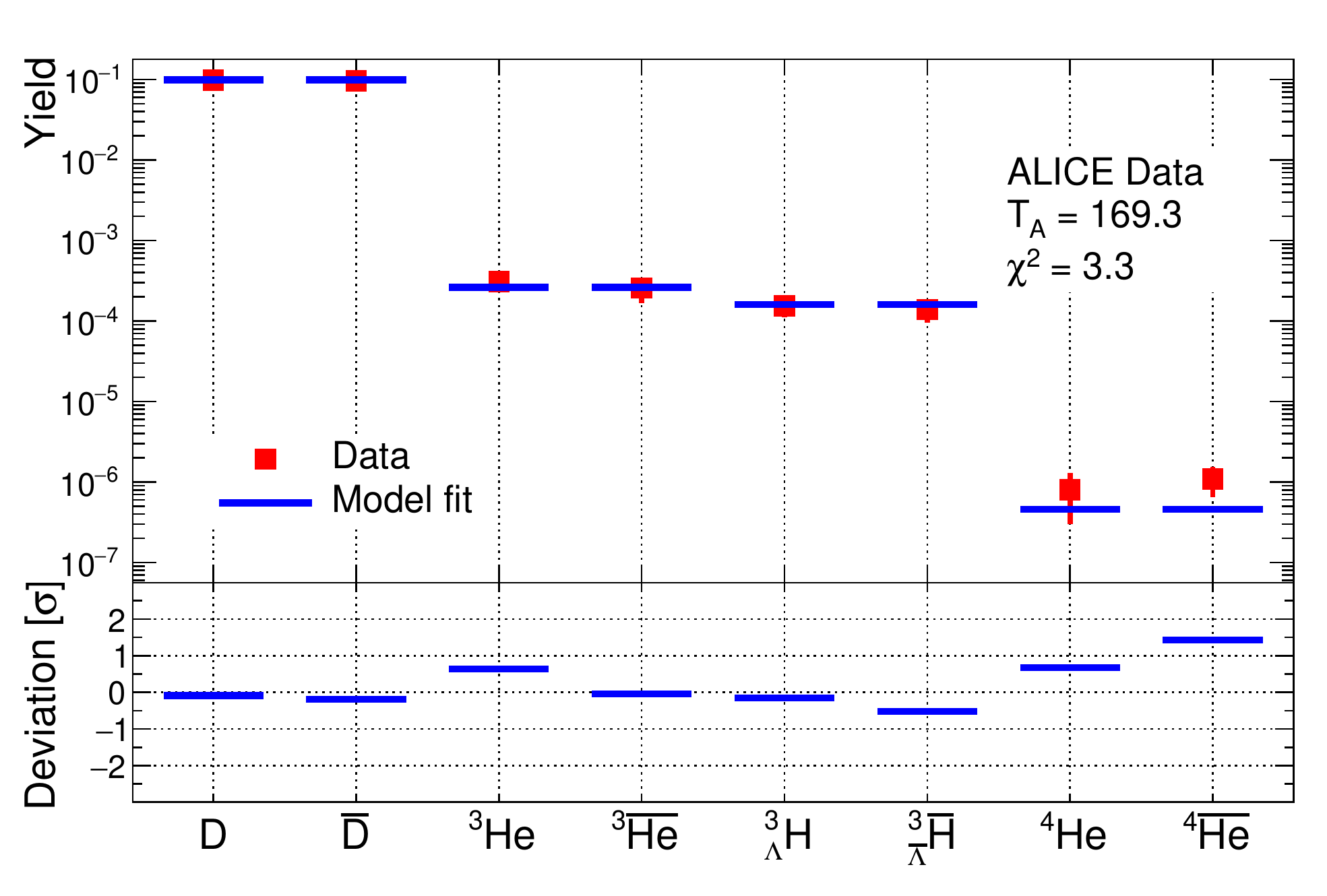}
	\includegraphics[width=0.5\columnwidth]{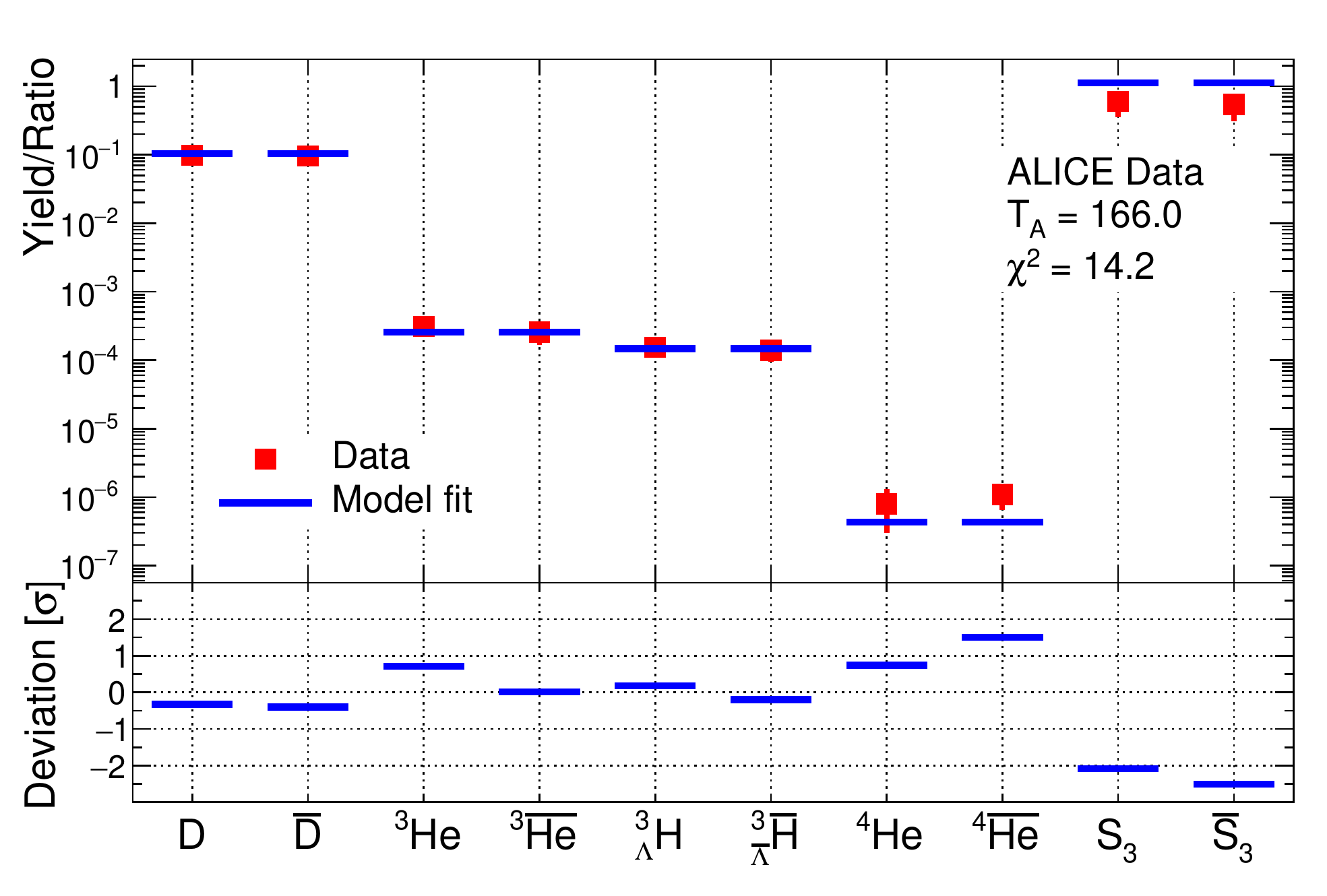}
	}
	\caption{ The yields of light nuclei measured at $\sqrt{s_{NN}} = 2.76$ TeV by ALICE vs. theoretical description by the IST EoS in the scenario II. Insertion shows the deviation of theory from data in the units of experimental error. 
		{\bf Left panel:} Separate CFO without fitting the $S_3$ and $\bar S_3$ ratios.
		{\bf Right panel:} Separate CFO with fitting the $S_3$ and $\bar S_3$ ratios.}
	\label{Fig1_KAB}
\end{figure}

For the scenario II (separate CFO of hadrons and light nuclei)  {there is} one additional parameter (three in total), i.e. the 
CFO temperature of nuclei $T_A$ since $T_A \neq T_h$ to fit either 18 (without fitting the $S_3$ and $\bar S_3$ ratios) or 20 (with fitting the $S_3$ and $\bar S_3$ ratios) data points. 

Fig.~\ref{Fig1_KAB} shows the results obtained for the IST EoS without fitting the $S_3$ and $\bar S_3$ ratios (left panel) and fitting them (right panel). Comparing these results, one can conclude that the inclusion of the $S_3$ and $\bar S_3$ ratios into a fit essentially worsens the quality of the data description. One can also see that the CFO temperature of nuclei $T_A$ decreases a bit 
 if the $S_3$ and $\bar S_3$ ratios are fitted. Note that the hadronic ratios are unchanged compared to the results of Refs. \citen{Ref1n,Ref2n}.
%
%
\begin{figure}[th]
	\centerline{\includegraphics[width=0.5\columnwidth]{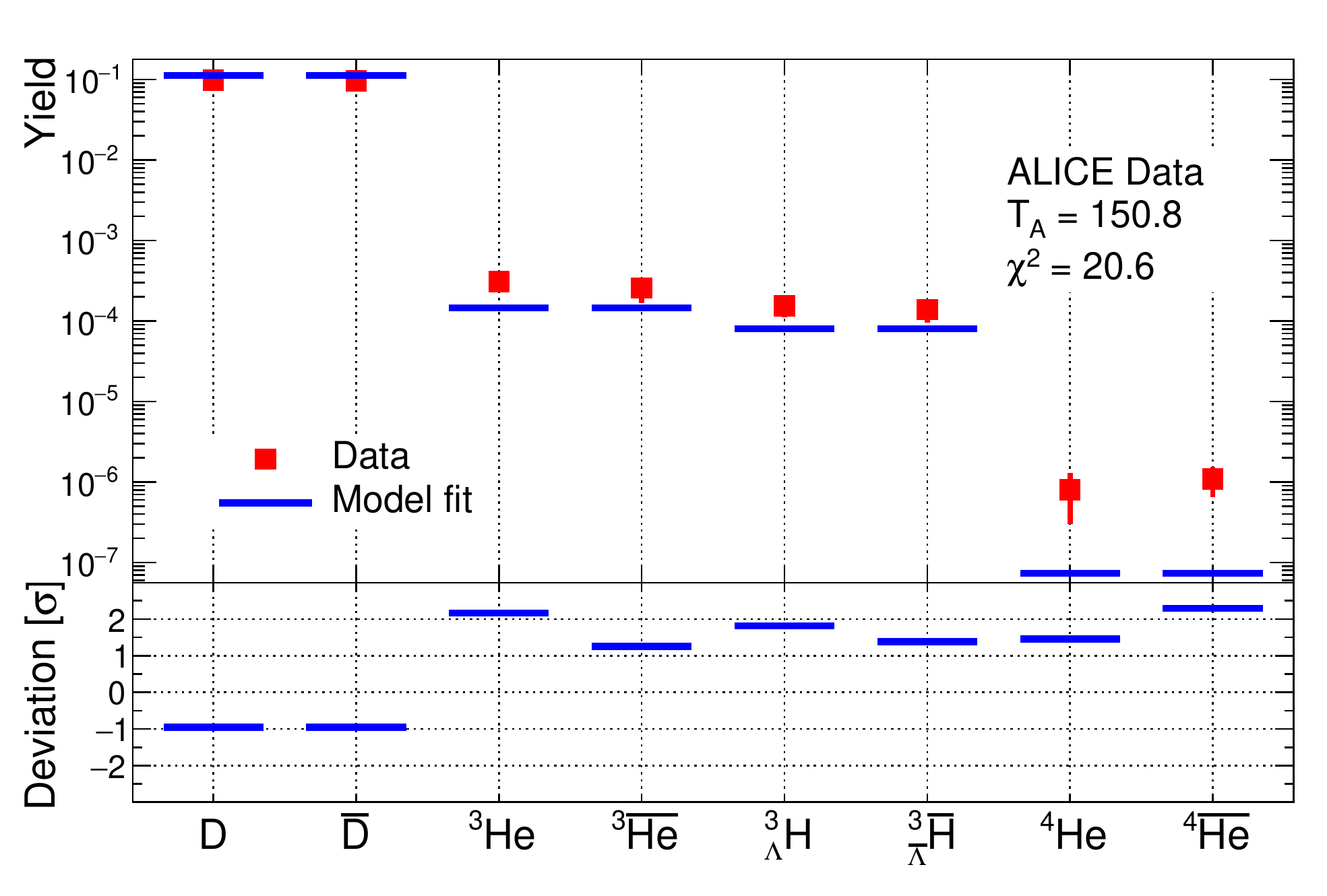}
	\includegraphics[width=0.5\columnwidth]{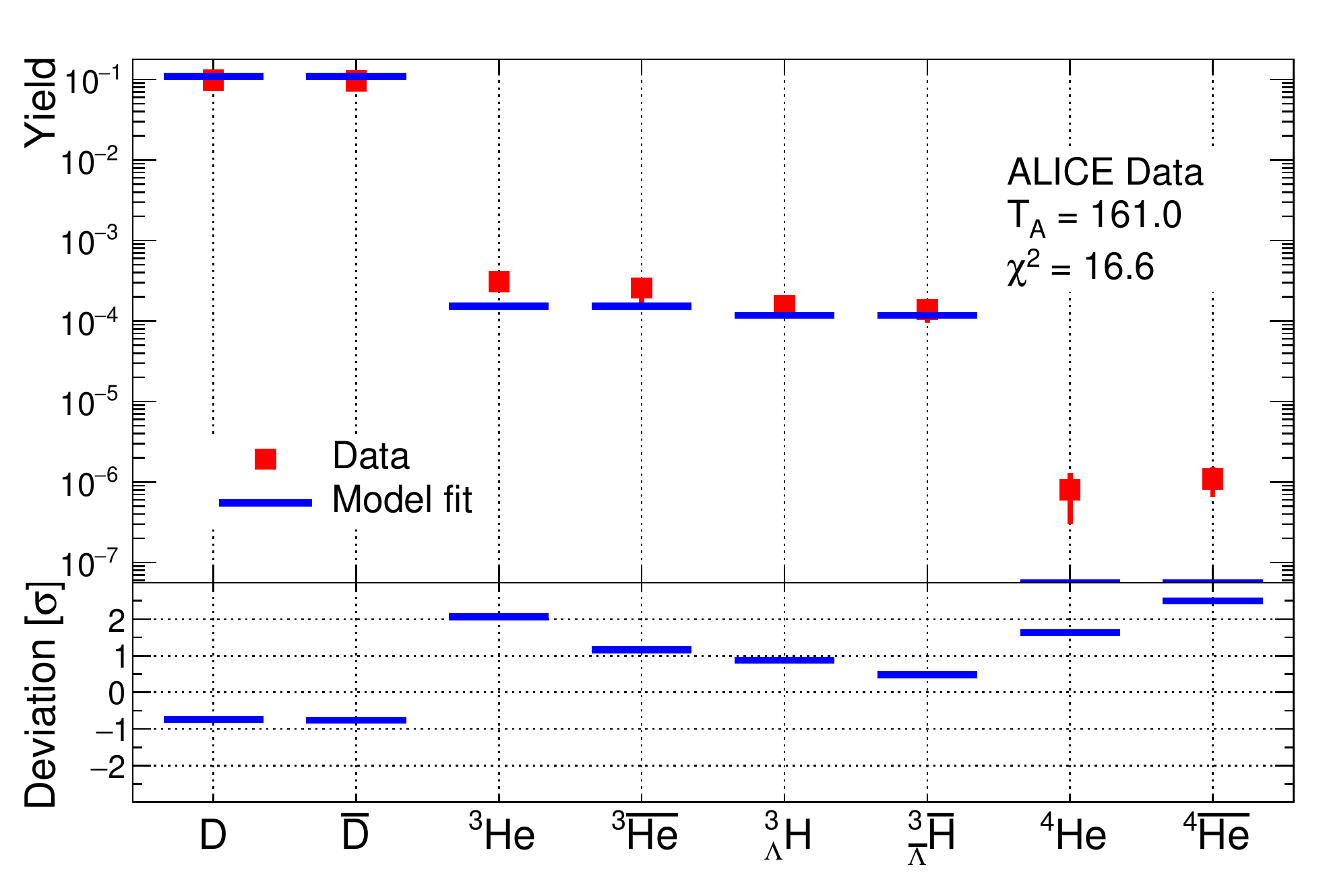}
	}
	\caption{The yields of light nuclei measured at $\sqrt{s_{NN}} = 2.76$ TeV by ALICE vs. theoretical description by the ISCT EoS in the scenario I (left panel) and in the scenario II (right panel) found without fitting the $S_3$ and $\bar S_3$ ratios.. Insertion shows the deviation of theory from data in the units of experimental error. 
}
	\label{Fig2_KAB}
\end{figure}

Fig.~\ref{Fig2_KAB} presents the results obtained for the ISCT EoS without fitting the $S_3$ and $\bar S_3$ ratios but for scenario I (left panel) and scenario II (right panel). From a comparison of these results, one can conclude that the light nucleus yields are very sensitive to the EoS! Note that overall worsening of the quality of the data description by ISCT EoS, compared to the ISC one, is not a great surprise to us, since the hard-core radii of hadrons used in our fit were found in Refs. \citen{IST2, IST3} exactly by the IST approach! Comparing the found values of $\chi^2/dof|_{I} \simeq 1.729$ for scenario I and $\chi^2/dof|_{II} \simeq 1.46$ for scenario
 II, one can conclude that scenario II is slightly more preferable.

\begin{figure}[th]
	\centerline{\includegraphics[width=0.5\columnwidth]{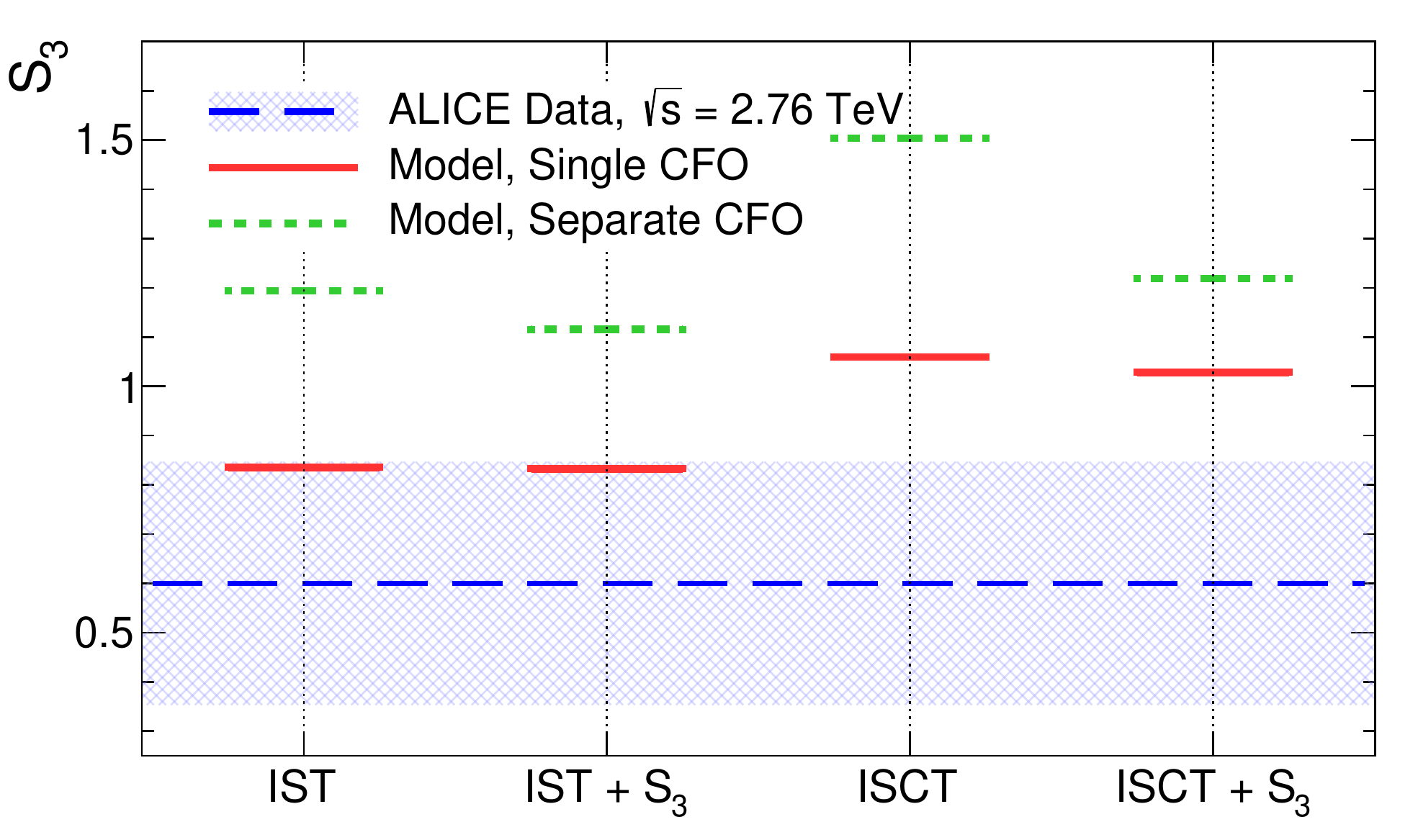}
	\includegraphics[width=0.5\columnwidth]{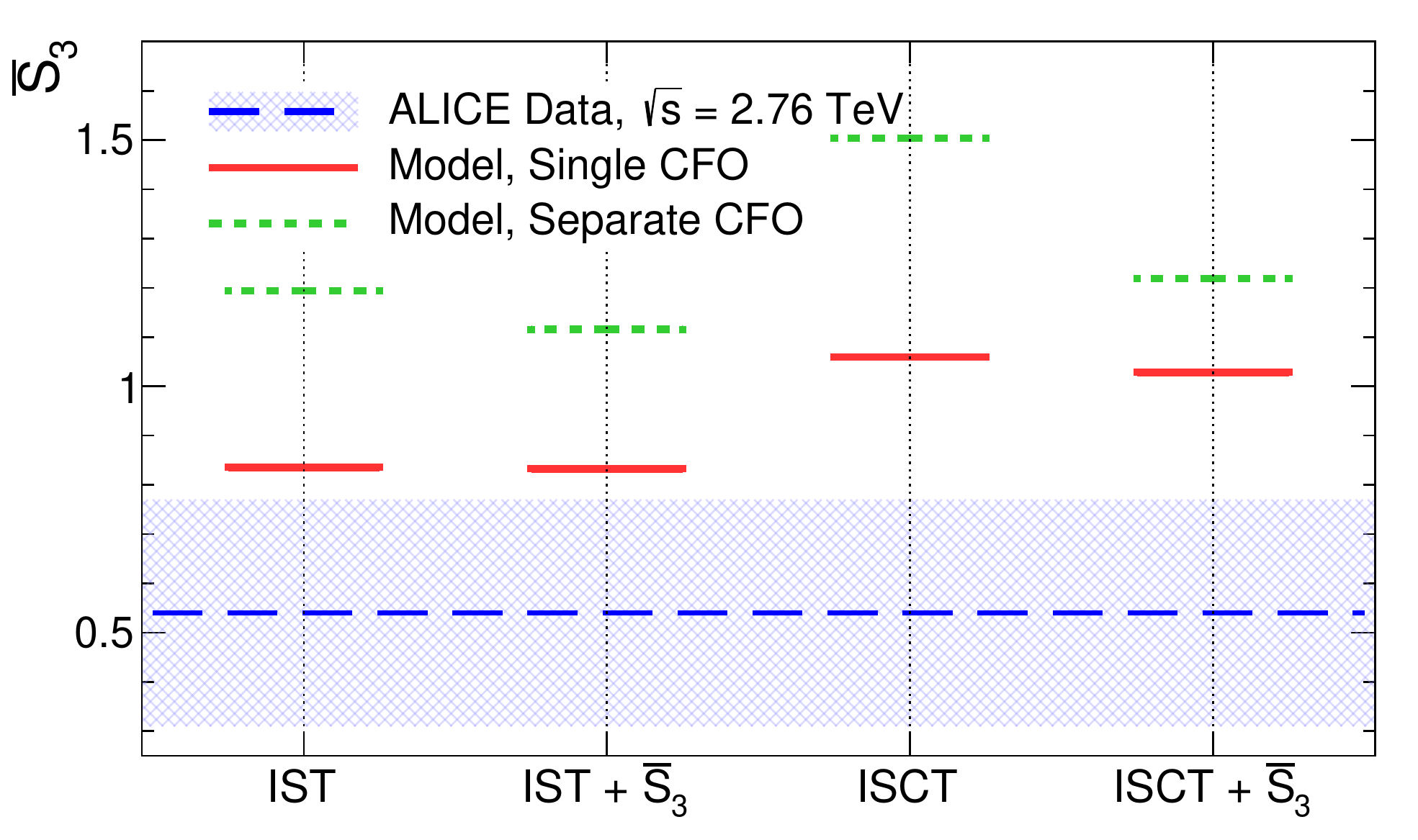}
	}
	\caption{ {\bf Left panel:} $S_3$ ratio measured at $\sqrt{s_{NN}} = 2.76$ TeV \cite{KAB_Ref1c}and its value obtained by the IST and ISCT EoS and without/with fitting the $S_3$ and $\bar S_3$ ratios.
		{\bf Right panel:} Same as in the left panel, but for $\bar S_3$.}
	\label{Fig3_KAB}
\end{figure}

The summary of results for the IST and ISCT EoS with/without fitting the $S_3$ and $\bar S_3$ ratios is given in Fig.~\ref{Fig3_KAB} and in Table 1. In our opinion, the results are inconclusive to a large extent and require further investigation. Although the minimal values of $\chi^2/dof$ are always achieved for scenario II (see Table 1), the obtained description of the $S_3$ and $\bar S_3$ ratios is always better for scenario I (see Fig.~\ref{Fig3_KAB}). Moreover, it seems that for the chosen set of hard-core radii of hadrons the IST EoS describes the data with higher quality. Of course, we could vary the hard-core radii of hadrons to get a better description of the data with the ISCT EoS, but, in our opinion, this is not the right way to improve the results. The right approach is to make a global fit of all existed data with the ISCT EoS and to determine the hard-core radii of hadrons that are suited for this EoS. However, this important task is far beyond the present work since the existing values of hadronic hard-core radii provide a rather satisfactory description of the data even with the ISCT EoS. 

\begin{table*}[!ht]
	{{\bf Table 1.} Results for the fit of ALICE data measured at $\sqrt{s_{NN}} = 2.76$ TeV obtained by the advanced HRGM. 
	The signs "-" or "+" in the second column indicate, respectively, that the $S_3$ and $\bar S_3$ ratios are ''not included" or 
	"included" in the fitting procedure. The CFO temperature of hadrons is $T_h$, the CFO temperature of light (anti)nuclei is $T_A$, while their CFO volume is $V_A$. The last column gives the fit quality. }
	\label{KAB_table1}
	\centering
	\begin{tabular}[t]{|l|c| c|c|c|c|}
\hline 
		Description & $S_3$ fit & $T_h,$ MeV & $T_A,$ MeV & $V_A,$ fm$^3$ & $\chi^2/dof$ \\ 

	\hline 	
		Scenario I, \, IST & - & $151.5 \pm 2.0$ & $151.5 \pm 2.0$ & $12227 \pm 2112$ & 1.33 \\ 
\hline 
		Scenario II, IST & - & $149.2 \pm 2.1$ & $169.3 \pm 5.6$ & $3906 \pm 1288$ & 0.716 \\ 
\hline 
	
		Scenario I, \, IST & + & $151.2 \pm 2.0$ & $151.2 \pm 2.0$ & $12521 \pm 2162$ & 1.323 \\ 
\hline 		
		Scenario II, IST & + & $151.5 \pm 2.0$ & $166.0 \pm 5.0$ & $4755 \pm 1471$ & 0.861 \\ 
\hline 		
		Scenario I, \, ISCT & - & $150.8 \pm 2.0$ & $150.8 \pm 2.0$ & $14428 \pm 2411$ & 1.729 \\ 
\hline 
		Scenario II, ISCT & - & $149.2 \pm 2.1$ & $161.0 \pm 4.6$ & $7683 \pm 2135$ & 1.460 \\ 
\hline 
	
		Scenario I, \, ISCT & + & $149.4 \pm 1.9$ & $149.4 \pm 1.9$ & $15882 \pm 2620$ & 1.968 \\ 
\hline 		
		Scenario II, ISCT & + & $149.2 \pm 2.1$ & $155.0 \pm 4.0$ & $1097 \pm 3001$ & 1.781 \\ 
\hline 		

	\end{tabular} 
\end{table*}

\section{Conclusions}

The present work is devoted to an investigation of the problem, which is traditionally hard for statistical mechanics, namely to a derivation and analysis of the grand canonical partition function of finite systems of interacting particles. {In our polemics} with the colleagues from other communities, this task is emerging again and again at least for the last twenty years when we are discussing the results of experiments on  collisions of heavy atomic nuclei.  A satisfactory and sufficiently rigorous solution of this problem is still required since it abuts into the development of a mathematically rigorous theory of phase transitions in finite systems. The present work, however, resolves this hard problem for the gas of Boltzmann particles with hard-core interaction in finite and even small systems of the volumes above 20 fm$^3$ with an idea to apply the obtained results to  modeling the properties of  the hadronic phase of QCD in a variety of experiments. 

The expressions derived for the IST EoS for finite volumes are analyzed in great details. One of the major results obtained here is as follows: it seems that the metastable states, which usually emerge in the finite systems with realistic interaction, can appear only at very high pressures at which the hadron resonance gas, most probably, is not applicable at all. Also,  it is shown how and under what conditions the obtained results for finite systems can be generalized in order to include into the suggested formalism the equation for curvature tension. The applicability range of the obtained equations of ISCT EoS for finite systems is discussed and their close relations to the equations of the morphological thermodynamics are established. It is necessary to stress that these relations established to the morphological thermodynamics give us a hope that the present results obtained just for the hard-core repulsion can be further generalized to more realistic and sophisticated interaction potentials of finite range. This is so since the concept of morphological thermodynamics is valid for the realistic potentials of finite range. 

Using the ISCT EoS we worked out a numerical code for the hadron resonance gas model and applied it to analyze the chemical freeze-out of hadrons and light nuclei with the number of (anti-)baryons not exceeding 4. The multiplicities of such particles were measured by the ALICE Collaboration in the central lead-lead collisions at the center-of-mass energy $\sqrt{s_{\rm NN}} =$ 2.76 TeV. 
A satisfactory description of these data can be obtained by the ISCT EoS, but the fit quality of the data achieved by the IST EoS is always higher. An apparent reason for this outcome is that the hard-core radii used in our simulations were previously obtained from the global fit of existing hadronic multiplicities with the hadron resonance gas model based exactly on the IST EoS. Interestingly, the yields of light nuclei turn out to be rather sensitive to the used EoS and, hence, the present analysis should be continued on a larger
set of data.

\vspace*{4.4mm}
\noindent
{\bf Acknowledgments.} 
The authors are thankful to Ivan Yakimenko for the valuable comments. 
KAB and GMZ acknowledge support from the NAS of Ukraine by its priority project "Fundamental properties of the matter in the relativistic collisions of nuclei and in the early Universe"
(No. 0120U100935).
VVS is thankful for the support by the Funda\c c\~ao para a Ci\^encia e Tecnologia (FCT), Portugal, by the project UID/04564/2021. 
The work of OII and DBB was supported by the Polish National Science Center (NCN) under grant No. 2019/33/B/ST9/03059. DBB received funding from the Russian Fund for Basic Research (RFBR) under grant No. 18-02-40137.
The work of LVB and EEZ was supported by the Norwegian Research Council (NFR) under grant No. 255253/ F53 CERN Heavy Ion Theory, and by the RFBR grants 18-02-40085 and 18-02-40084. 
KAB, OVV, NSYa and LVB thank the Norwegian Agency for International Cooperation and Quality Enhancement in Higher Education for the financial support under grants CPEA-LT-2016/10094 and UTF-2016-long-term/10076.
The research of AVT was funded by the RFBR under the grant No. 18-02-40086 and partially supported by the Ministry of Science and Higher Education of the Russian Federation, Project ``Fundamental properties of elementary particles and cosmology" 
No 0723-2020-0041.
The authors are grateful to the COST Action CA15213 ``THOR" for supporting their networking.

\bibliographystyle{ref_style}
\bibliography{bibliography}

\end{document}